\documentclass[12pt,twoside,fleqn]{article}
\usepackage{epsf,espcrc1,rotating}
\begin{document}
\title{
{\it \small To be published in:  Recent Developments and Applications of
Density Functional Theory, ed.~J.~M.~Seminario, (Elsevier, Amsterdam, 1996)
\vskip 6mm}
\bf Generalized gradient approximations to density functional theory:
comparison with exact results}

\author{Claudia Filippi\address{Laboratory of Atomic and Solid State Physics,
Cornell University, Ithaca, NY 14853},
Xavier Gonze\address{Unit\'e P.C.P.M., Universit\'e Catholique de Louvain,
B-1348 Louvain-la-Neuve, Belgium} and
C. J. Umrigar\address{Cornell Theory Center, Cornell University, Ithaca,
NY 14853}$^{,\,{\rm a}}$}

\maketitle
\begin{abstract}
In order to assess the accuracy of commonly used approximate
exchange-correlation density functionals, we present a comparison of
accurate exchange and correlation potentials, exchange energy densities
and energy components with the corresponding approximate quantities.
Four systems are used as illustrative examples: the model system of two
electrons in a harmonic potential and the He, Be and Ne atoms.
A new ingredient in the paper is the separation of the exchange-correlation
potential into exchange and correlation according to the density functional
theory definition.
\end{abstract}

\section{INTRODUCTION}
\label{s0}

Within density functional theory (DFT), the ground state energy of an
interacting system of electrons in an external potential can be written as
a functional of the ground state electronic density~\cite{HK}.
When compared to conventional quantum chemistry methods, this approach is
particularly appealing since it does not rely on the knowledge of the complete
{\it N}-electron wave function but only of the electronic density.
Unfortunately, although the theory is in principle exact, the energy functional
contains an unknown quantity, called the exchange-correlation energy,
$E_{\rm xc}\left[\rho\right]$, that must be approximated in any practical
implementation of the method.
Several approximate exchange-correlation functionals have been proposed in the
literature, the most commonly used ones being the local density approximation
(LDA) and the generalized gradient approximation (GGA).

The local density approximation~\cite{KS} is the simplest and most widely used
exchange-cor\-relation functional:
\begin{eqnarray}
E^{\rm LDA}_{\rm xc}\left[\rho\right]=\int \rho({\bf r})\,
\epsilon^{\rm LDA}_{\rm xc}(\rho({\bf r}))\,{\rm d}{\bf r},
\end{eqnarray}
where $\epsilon^{\rm LDA}_{\rm xc}(\rho)$ is the exchange-correlation
energy per particle of a homogeneous electron gas of density $\rho$.
$E^{\rm LDA}_{\rm xc}\left[\rho\right]$ is by definition the exact functional
for a homogeneous electron gas and has been shown to give also a
qualitatively good description of the ground state properties of a variety of
highly inhomogeneous systems~\cite{JG}.
However, LDA does not always provide sufficiently accurate results.
For example, it almost always overestimates the binding energy and
underestimates the bond-length of weakly bound molecules and solids~\cite{JG}.
LDA fails also to predict the ground state structure of iron, although the
error is quantitatively small~\cite{WKK}.

In an attempt to improve upon LDA, a dependence of the exchange-correlation
energy on the derivatives of the electronic density can be introduced.
A simple and systematic extension of LDA is the gradient expansion
approximation (GEA)~\cite{KS},
\begin{eqnarray}
E^{\rm GEA}_{\rm xc}[\rho]= \int \rho({\bf r})
\epsilon^{\rm LDA}_{\rm xc}(\rho({\bf r})) \,{\rm d}{\bf r}
+ \int B_{\rm xc}(\rho({\bf r})) | \nabla \rho ({\bf r})|^2 \,
{\rm d}{\bf r} + \cdots,
\label{GEA}
\end{eqnarray}
which is is asymptotically valid for densities that vary slowly over space.
For finite systems, the terms in the expansion of the exchange energy of order
greater or equal than four diverge while the exchange potential diverges
already at second order.
By using a constant $B_{\rm xc}$ determined variationally for each atom,
Herman {\it et al.} showed that the gradient expansion up to second order
yields improved total energies of atoms~\cite{Herman69}.
An improvement in the exchange energies is also obtained when the exact
expansion coefficient for exchange, $B_{\rm x}$, is used in an exchange-only
calculation~\cite{PK}. However, the inclusion of correlation in the second
order expansion gives energies that are less accurate than LDA~\cite{MB,PLS}.

The analysis of this failure has lead some authors~\cite{LM83,Perdew85} to
introduce a generalization of the gradient expansion, whose generic functional
form (here restricted to second-order derivative) is
\begin{eqnarray}
E^{\rm GGA}_{\rm xc}\left[\rho\right]=\int\, \rho({\bf r})\,
\epsilon^{\rm GGA}_{\rm xc}(\rho({\bf r}),\left|\nabla\rho({\bf r})\right|,
\nabla^2\rho({\bf r}))\,{\rm d}{\bf r}. \label{gga}
\end{eqnarray}
Many different GGA's have been proposed since
then~\cite{LM83,B88,PW86X,PW86C,PW91,PW92,WL,LYP,ECMV} but none of them is
clearly better than the others.
Other classes of approximate functionals have also been tested, including
schemes that remove the self-interaction of electrons~\cite{SIC,OEP},
the weighted density approximation~\cite{GJL,GJ} and the average density
approximation~\cite{GJL} that attempt to model the exchange-correlation hole.
However, generalized gradient approximations have recently received particular
attention both in the physics and chemistry communities since they do not
add appreciable computational complexity to the LDA scheme.

The GGA functionals seem to give a good description of several finite systems:
they significantly improve total energies of atoms~\cite{P}, as well as binding
energies~\cite{P,KP,B92,JGP} and vibrational frequencies~\cite{KP} of first
and second-row molecules.
They also give better estimates of bond-lengths and binding energies of
weakly bound systems such as IIA and IIB homonuclear dimers~\cite{OB} and
hydrogen bonded systems~\cite{WATER}, which are greatly overbound in LDA.
Lattice constants, bulk moduli and cohesive energies of simple metals~\cite{P}
and 3{\it d} transition metals~\cite{BJG} are also improved.
For iron, the correct ferromagnetic bcc ground state is predicted~\cite{BJG}.
On the other hand, the lattice constants and the bulk moduli of the
semiconductors Si, Ge and GaAs are less accurately predicted by the GGA's than
LDA, the effect being particularly large for the bulk moduli~\cite{semicond}.
Calculations for 4{\it d} and 5{\it d} metals indicate that, while LDA tends
to slightly underestimate lattice constants and overestimate bulk moduli,
GGA's often overcorrect, sometimes leading to results in worse agreement with
experiment~\cite{metal}.
The tendency of the GGA's to yield smaller binding energies and larger
bond-lengths than LDA is explained by the fact that the GGA's favor
inhomogeneity in the density.
Since isolated atoms are more inhomogeneous than molecules or solids, the
energy of the atoms is lowered more, resulting in smaller binding energies.
Similarly, for weakly bound systems, larger bond-lengths lead to increased
inhomogeneity and are therefore favored by the GGA's.

Studies that compare the predicted structural and energetic properties of
various systems with the corresponding experimental quantities add very little
to the understanding of the reasons for the successes or the failures of the
approximate functionals.
In order to gain more insight into the limitations of the approximate
functionals, many
researchers~\cite{S79,vonBarth84,AP,AS,NM,D90,Chen94,LB1,MZ,Ludena95,BBS,GLB,IH,proc,KG,UG}
have compared accurate density functional properties, calculated by more
sophisticated and more computationally demanding methods, with the
corresponding density functional properties calculated from the approximate
functionals.
In this review, we limit ourselves to some of our earlier work and discuss
four systems, for which it is possible to obtain exact or very accurate
wave functions, as illustrative examples: a model system of two electrons 
in a harmonic potential and the He, Be and Ne atoms.
While, we mostly concentrate on the description of our previous work,
there is a new ingredient in this paper: the separation of the
exchange-correlation potential of many-electron systems into separate exchange
and correlation components according to the DFT definition.

The outline of the rest of the paper is as follows.  In Sec.~\ref{s1}, we
briefly introduce density functional theory and its Kohn-Sham formulation.
In Sec.~\ref{s2}, we list some properties of the exact exchange-correlation
functional and we determine which properties are satisfied or violated by
LDA and various GGA's. The short and long distance asymptotic behavior of the
exchange-correlation potential and local exchange energy per electron is
also discussed. In Sec.~\ref{s3}, we derive the formulae used to determine
the exact density functional quantities and, in Sec.~\ref{s4}, we describe the
systems we study and the corresponding densities.
The exact exchange and correlation potentials, local exchange energies per
electron and components of the total energy of these systems are compared with
the corresponding approximate LDA and GGA quantities in Secs.~\ref{s5},
\ref{s6} and~\ref{s7}, respectively. In Sec.~\ref{s8}, we summarize our
conclusions and discuss some prospects for better approximate functionals.
In Appendix~\ref{a1}, we list the approximate functionals used in the
comparison.

\section{THEORETICAL BACKGROUND}
\label{s1}

Density functional theory provides an expression for the ground
state energy of a system of interacting electrons in an external potential
as a functional of the ground state electronic density \cite{HK}.
Let us assume for simplicity that the spin polarization of the system
of interest is identically zero.
In the Kohn-Sham formulation of density functional theory \cite{KS}, the ground
state density is written in terms of single-particle orbitals obeying the
equations in atomic units ($\hbar=e=m=1$):
\begin{eqnarray}
\left\{-\frac{1}{2}\nabla^2+v_{\rm ext}({\bf r})+\int \frac{\rho({\bf r}')}
{\left|{\bf r}-{\bf r}'\right|}{\rm d}{\bf r}'+v_{\rm xc}\left(
\left[\rho\right];{\bf r}\right)\right\}\psi_i=\epsilon_i\psi_i,
\label{KS}
\end{eqnarray}
where
\begin{eqnarray}
\rho({\bf r})=\sum_{i=1}^N\left|\psi_i({\bf r})\right|^2.
\label{rho}
\end{eqnarray}
The electronic density is constructed by summing over the {\it N} lowest
energy orbitals where {\it N} is the number of electrons.
$v_{\rm ext}({\bf r})$ is the external potential.
The exchange-correlation potential $v_{\rm xc}\left(\left[\rho\right];
{\bf r}\right)$ is the functional derivative of the exchange-correlation
energy $E_{\rm xc}\left[\rho\right]$ that enters in the expression for
the total energy of the system:
\begin{eqnarray}
E=-\frac{1}{2}
\sum_{i=1}^N\int\psi_i\nabla^2\psi_i\,{\rm d}{\bf r}
+\int\,\rho\left({\bf r}\right)v_{\rm ext}
\left({\bf r}\right)\,{\rm d}{\bf r}
+\frac{1}{2}\int\!\!\int\frac{\rho({\bf r})
\rho({\bf r}')}{\left|{\bf r}-{\bf r}'\right|} {\rm d}{\bf r}\,{\rm d}{\bf r}'
+E_{\rm xc}\left[\rho\right].
\label{eq0}
\end{eqnarray}
The exchange-correlation functional is written as the sum of two separate
contributions for exchange and correlation,
\begin{eqnarray}
E_{\rm xc}\left[\rho\right]=E_{\rm x}\left[\rho\right]+
E_{\rm c}\left[\rho\right].\label{separaE}
\end{eqnarray}
The definition of the exchange energy is in terms of the non-interacting
wave function $\Phi_0$, the Slater determinant constructed from the Kohn-Sham
orbitals, as
\begin{eqnarray}
E_{\rm x}\left[\rho\right]=
\left<\Phi_0\right|V_{\rm ee}\left|\,\Phi_0\right>-
\frac{1}{2}\int\!\!\int\frac{\rho({\bf r})
\rho({\bf r}')}{\left|{\bf r}-{\bf r}'\right|}
{\rm d}{\bf r}\,{\rm d}{\bf r}',\label{enx0}
\end{eqnarray}
where $V_{\rm ee}$ is the electron-electron interaction.
This definition differs from the conventional quantum chemistry definition of
$E_{\rm x}$ as the exchange energy in a Hartree-Fock calculation, given
by the same expression as in Eq.~\ref{enx0} but with the Kohn-Sham
determinant replaced by the Hartree-Fock determinant.
The separation of the exchange-correlation energy functional into the 
separate exchange and correlation components yields a corresponding 
splitting of the exchange-correlation potential into 
$v_{\rm x}\left(\left[\rho\right];{\bf r}\right)$ and
$v_{\rm c}\left(\left[\rho\right];{\bf r}\right)$.
In this formulation, the essential unknown quantity is the exchange-correlation
energy $E_{\rm xc}\left[\rho\right]$. If the functional form of $E_{\rm xc}
\left[\rho\right]$, and consequently the exchange-cor\-re\-la\-tion potential,
were available, we could solve the {\it N}-electron problem by finding the
solution of a set of single-particle equations.
However, as mentioned in Sec.~\ref{s0}, the exact functional form of
$E_{\rm xc}\left[\rho\right]$
is not known and it is necessary to make approximations for this term.

\section{PROPERTIES OF THE EXACT DENSITY FUNCTIONAL}
\label{s2}

In order to understand if it is possible to construct an approximate
functional that does not suffer from the weaknesses of the existing GGA's,
it is useful to study the known properties of the exact density functional
and which of these are violated by the commonly used approximate functionals.
There are actually a surprisingly large number of known
properties~\cite{GLxc,LP93,LP85,L91,GLs,LPr} and some of these are listed in
Table~\ref{knownprop}.
The GGA functionals compared in the table can be found in Appendix~\ref{a1}.

\begin{table}[tbp]
\caption[]{Some properties of the exact exchange-correlation functional.
Comparison with approximate functionals. Taken from Ref.~\cite{proc} and
modified.}
\label{knownprop}
\small
\renewcommand{\arraystretch}{1.3}
\begin{tabular*}{\textwidth}{@{}l@{\extracolsep{\fill}}ccccccc}
\hline
\hspace{.7cm}{\normalsize Property}
 & $E^{\rm LDA}_{\rm xc}$
 & $E^{\rm LM}_{\rm xc}$
 & $E^{\rm PW91}_{\rm xc}$
 & $E^{\rm B88}_{\rm x}$
 & $E^{\rm ECMV}_{\rm x}$
 & $E^{\rm WL}_{\rm c}$
 & $E^{\rm LYP}_{\rm c}$ \\
 & \cite{PW92}
 & \cite{LM83}
 & \cite{PW91}
 & \cite{B88}
 & \cite{ECMV}
 & \cite{WL}
 & \cite{LYP}\\
\hline
$\,\,\,$1 \ \
$\rho_{\rm x}({\bf r},{\bf r}')\leq0$
&  {\small Y}  & --- &  {\small Y}  & --- & --- & --- & --- \\
$\,\,\,$2 \ \
$\int \rho_{\rm x}({\bf r},{\bf r}')\,{\rm d}{\bf r}'=-1$
&  {\small Y}  & --- &  {\small Y}  & --- & --- & --- & --- \\
$\,\,\,$3 \ \
$\int \rho_{\rm c}({\bf r},{\bf r}')\,{\rm d}{\bf r}'=0$
&  {\small Y}  & --- &  {\small Y}  & --- & --- & --- & --- \\
\hline
$\,\,\,$4 \ \
$E_{\rm x}\left[\rho\right]<0$
&  {\small Y}  &  {\small Y}  &  {\small Y}  &  {\small Y}  &  {\small Y}  & --- & --- \\
$\,\,\,$5 \ \
$E_{\rm c}\left[\rho\right]\leq 0$
&  {\small Y}  &  {\small N}  &  {\small N}  & --- & --- &  {\small N}  &  {\small N}  \\
$\,\,\,$6 \ \
$E_{\rm x}\left[\rho\right], E_{\rm xc}\left[\rho\right]\geq -c\int \rho^{4/3}\,{\rm d}{\bf r}\;\;^a$
&  {\small Y}  &  {\small N}  &  {\small Y}  &  {\small N}  &  {\small N}  & --- & --- \\
\hline
$\,\,\,$7 \ \
$E_{\rm x}\left[\rho_\lambda\right]=\lambda E_{\rm x}\left[\rho\right]$
&  {\small Y}  &  {\small Y}  &  {\small Y}  &  {\small Y}  &  {\small Y}  & --- & --- \\
$\,\,\,$8 \ \
$E_{\rm c}\left[\rho_\lambda\right]<\lambda E_{\rm c}\left[\rho\right],\;\lambda<1\;\;^b$
&  {\small Y}  &  {\small N}  &  {\small Y}  & --- & --- &  {\small N}  &  {\small N}  \\
$\,\,\,$9 \ \
$\lim_{\lambda\rightarrow \infty} E_{\rm c}\left[\rho_\lambda\right]> -\infty$
&  {\small N}  &  \ \ {\small Y}$\;^c$ &  \ \ {\small Y}$\;^c$ & --- & --- &  {\small Y}  &  {\small Y}  \\
10 \ \
$\lim_{\lambda\rightarrow 0} \frac{1}{\lambda}
 E_{\rm c}\left[\rho_\lambda\right]>-\infty$
& {\small Y} &  {\small N}  &  {\small Y}  & --- & --- &  {\small Y}  &  {\small Y}  \\
\hline
11 \ \
$\lim_{\lambda\rightarrow\infty}
 E_{\rm x}\left[\rho^{\rm x}_{\lambda}\right]>-\infty$
&  {\small N}  &  {\small N}  &  {\small Y}  &  {\small N}  &  {\small N}  & --- & --- \\
12 \ \
$\lim_{\lambda\rightarrow 0}
 E_{\rm x}\left[\rho^{\rm x}_{\lambda}\right]>-\infty$
&  {\small Y}  &  {\small N}  &  {\small Y}  &  {\small Y}  &  {\small Y}  & --- & --- \\
13 \ \
$\lim_{\lambda\rightarrow \infty} \frac{1}{\lambda}
 E_{\rm x}\left[\rho^{\rm xy}_\lambda\right]>-\infty$
&  {\small Y}  &  {\small N}  &  {\small Y}  &  {\small Y}  &  {\small Y}  & --- & --- \\
14 \ \
$\lim_{\lambda\rightarrow 0} \frac{1}{\lambda}
 E_{\rm x}\left[\rho^{\rm xy}_{\lambda \lambda}\right]>-\infty\;\;$
&  {\small N}  &  {\small N}  &  {\small Y}  &  {\small N}  &  {\small N}  & --- & --- \\
15 \ \
$\lim_{\lambda\rightarrow\infty}
 \lambda E_{\rm c}\left[\rho^{\rm x}_{\lambda}\right]>-\infty$
&  {\small N}  &  \ \ {\small Y}$\;^c$ &  {\small Y}  & --- & --- &  {\small N}  &  {\small N}  \\
16 \ \
$\lim_{\lambda\rightarrow 0} \frac{1}{\lambda}
 E_{\rm c}\left[\rho^{\rm x}_{\lambda}\right]=0\;\;$
&  {\small N}  &  {\small N}  &  {\small Y}  & --- & --- &  {\small N}  &  {\small N}  \\
17 \ \
$\lim_{\lambda\rightarrow \infty}
 E_{\rm c}\left[\rho^{\rm xy}_{\lambda \lambda}\right]=0\;\;$
&  {\small N}  &  {\small N}  &  {\small Y}  & --- & --- &  {\small N}  &  {\small N}  \\
18 \ \
$\lim_{\lambda\rightarrow 0}
\frac{1}{\lambda^2} E_{\rm c}\left[\rho^{\rm xy}_{\lambda \lambda}\right]>-\infty\;\;$
&  {\small N}  &  \ \ {\small Y}$\;^c$ &  {\small Y}  & --- & --- &  \ \ {\small Y}$\;^c$  &  {\small N}  \\
\hline
19 \ \
$\epsilon_{\rm x}(r)\rightarrow -1/2r,\;\;r \rightarrow\infty$
&  {\small N}  &  {\small N}  &  {\small N}  &  \ {\small YN}$\;^d$ &  {\small N}  & --- & --- \\
20 \ \
$v_{\rm x}(r)\rightarrow -1/r,\;\;r \rightarrow\infty$
&  {\small N}  &  {\small N}  &  {\small N}  &  {\small N}  &  {\small N}  & --- & --- \\
21 \ \
$v_{\rm x}(r), v_{\rm c}(r)\,{\rm finite},\;r \rightarrow 0$ \
&  {\small Y}  &  {\small N}  &  {\small N}  &  {\small N}  &  {\small N}  &  {\small N}  &  {\small N}  \\
\hline
22 \ \
LDA limit for constant $\rho$
&  {\small Y}  &  {\small N}  &  {\small Y}  &  {\small Y}  &  {\small Y}  &  {\small N}  &  {\small N}  \\
23 \ \
GEA limit for slowly varying $\rho$
& --- &  {\small N}  &  {\small Y}  &  {\small N}  &  {\small N}  &  {\small N}  &  {\small N}  \\
\hline
\multicolumn{8}{@{}l}{$^a\, 1.44<c<1.68$}\\
\multicolumn{8}{@{}l}{$^b\,$ Note that
$E_{\rm c}\left[\rho_\lambda\right]<\lambda E_{\rm c}\left[\rho\right],\;
\lambda<1$ is equivalent to
$E_{\rm c}\left[\rho_\lambda\right]>\lambda E_{\rm c}\left[\rho\right],\;
\lambda>1$.}\\
\multicolumn{8}{@{}l}{$^c\,$ But it diverges to $+\infty$. The PW91 GGA
can be modified to satisfy this relation~\cite{LP93}.}\\
\multicolumn{8}{@{}l}{$^d$\, ``{\small Y}'' for exponential $ \rho({\bf r})$
but ``{\small N}''~in general, {\it e.g.}\ $\epsilon_{\rm x}^{\rm B88}(r)
\rightarrow-1/r$ for a Gaussian density.}
\end{tabular*}
\end{table}
The first group of properties in Table~\ref{knownprop} are sum rules that
are satisfied by the exchange and correlation holes.
The exchange-correlation energy functionals can be interpreted as the energy
arising form the interaction of an electron at ${\bf r}$ and its
exchange-correlation hole at ${\bf r}'$,
\begin{eqnarray}
E_{\rm xc}\left[\rho\right]=\frac{1}{2}\int\!\!\int\,\frac{\rho
\left({\bf r}\right)\rho_{\rm xc}\left({\bf r},{\bf r}'\right)}
{\left|{\bf r}-{\bf r}'\right|}\,{\rm d}{\bf r}\,{\rm d}{\bf r}'.
\end{eqnarray}
The exchange-correlation hole is then separated into exchange and correlation,
$\rho_{\rm x}({\bf r},{\bf r}')$ and $\rho_{\rm c} ({\bf r},{\bf r}')$
respectively, where the exchange contribution comes from the non-interacting
system according to Eq.~\ref{enx0}.
The exchange-correlation hole can be expressed as an integral over the coupling
constant of the density-density correlation function of the interacting
system~\cite{GLxc}.

The second group of properties provide bounds on the functionals.
Condition 6 is known as the Lieb-Oxford bound and a tighter version of it
is given in Ref.~\cite{LP93}:
\begin{eqnarray}
\lim_{\lambda\rightarrow 0}\frac{1}{\lambda}E_{\rm xc}\left[\rho_\lambda\right]
=\inf_{\Psi\rightarrow\rho}\left<\Psi\left|V_{\rm ee}\right|\Psi\right>-
\frac{1}{2}\int\!\!\int\frac{\rho({\bf r})
\rho({\bf r}')}{\left|{\bf r}-{\bf r}'\right|}
{\rm d}{\bf r}\,{\rm d}{\bf r}'
\ge -c\int \rho({\bf r})^{4/3}\,{\rm d}{\bf r},
\end{eqnarray}
where the density $\rho_\lambda$ is obtained by uniformly scaling the
density $\rho$ in all three spatial directions:
\begin{eqnarray}
\rho_\lambda({\bf r})=\lambda^3 \rho\left(\lambda{\bf r}\right).
\label{scale}
\end{eqnarray}
This scaled density integrates to the same number of electrons as the
unscaled one.

The scaling of the exchange and correlation functionals when the charge
density is scaled uniformly~\cite{LP85,L91} is described by the third group of
properties.
It now becomes evident why it is useful to separate the exchange component
from the entire functional according to Eq.~\ref{enx0}. For exchange, an exact
relation (condition 7) exists under uniform density scaling that determines
how derivatives of the density combine with the density in an exchange GGA
functional:
\begin{eqnarray}
E^{\rm GGA}_{\rm x}\left[\rho\right]=\int\,\rho({\bf r})^{4/3}\,
F(\,\left|\nabla\rho({\bf r})\right|/\rho({\bf r})^{4/3},
\nabla^2\rho({\bf r})/\rho({\bf r})^{5/3},\ldots\,)\,{\rm d}{\bf r}.
\end{eqnarray}

The fourth set of properties consists of the relations under non-uniform
scaling of the density in one or two of the spatial directions:
\begin{eqnarray}
\rho^{\rm x}_\lambda({\bf r})=\lambda\,\rho(\lambda x, y, z),\;\;\;
\rho^{\rm xy}_{\lambda\lambda}({\bf r})=\lambda^2\,\rho\left(\lambda x,
\lambda y,z\right).
\end{eqnarray}
The non-uniform scaling relations for the correlation energy are here given in
the tightest form and are derived under the assumption of the existence of
a Taylor series in $\lambda$ or $1/\lambda$~\cite{GLs}.
A summary of the scaling relations and their derivation is provided in
Ref.~\cite{LPr}.

The fifth group of properties describes the long-distance asymptotic
behavior (for finite systems) of the exchange potential and the local exchange
energy per electron (see Sec.~\ref{s2}) and the short-distance behavior of the
exchange and correlation potential.

The last two conditions are that the functionals must reduce to the LDA
functional in the limit of a  homogeneous density and to the correct
second-order expansion for a slowly varying density.

We also mention an additional relation not included in the table,
the convexity constraint~\cite{LP93}, which is given by
\begin{eqnarray}
\left.\frac{\partial^2B\left[\rho+\epsilon\Delta\rho\right]}{\partial\epsilon^2}
\right|_{\epsilon=0}+\int\!\!\int\frac{\Delta\rho({\bf r})\Delta\rho({\bf r}')}
{\left|{\bf r}-{\bf r}'\right|} {\rm d}{\bf r}\,{\rm d}{\bf r}'\ge 0,
\end{eqnarray}
for arbitrary $\Delta\rho$ integrating to zero, where
\begin{eqnarray}
B\left[\rho\right]=\lim_{\lambda\rightarrow 0}
\frac{1}{\lambda}E_{\rm xc}\left[\rho_\lambda\right].
\end{eqnarray}
This constraint is very stringent: it is violated by the LDA, by the
Perdew-Wang '91 GGA and, likely, by any other GGA functional~\cite{LP93}.

From Table~\ref{knownprop}, it appears that most of the functionals violate
the asymptotic behavior of the exchange-correlation potential and local exchange
energy per electron.
Within GGA, the exchange-correlation potential is given by the functional
derivative of $E_{\rm xc}^{\rm GGA}\left[\rho\right]$ (Eq.~\ref{gga}):
\begin{eqnarray}
v_{\rm xc}\left(\left[\rho\right];{\bf r}\right)=\left[\frac{\partial\,
e_{\rm xc}} {\partial\,\rho}-\nabla\cdot\left(\frac{\partial\,
e_{\rm xc}}{\partial\,\nabla \rho} \right)+
\nabla^2\left(\frac{\partial\,e_{\rm xc}}
{\partial\,\nabla^2 \rho}\right)\right]_{\rho({\bf\footnotesize  r}),
\nabla\rho({\bf\footnotesize r}),\nabla^2\rho({\bf\footnotesize r})},
\label{pot}
\end{eqnarray}
where $e_{\rm xc}$ is the exchange-correlation energy density, $e_{\rm xc}=\rho
\,\epsilon_{\rm xc}$. As pointed out in our earlier papers~\cite{harmonium,he},
any GGA that has terms containing $\nabla\rho$ but no higher derivatives
of $\rho$ must yield an exchange-correlation potential that diverges at
nuclei. If $\rho\sim{\rm constant}+r^s$ near the origin, then
$\left|\nabla\rho\right|\sim r^{s-1}$.
If the leading behavior of $e_{\rm xc}$ in the gradient, near an extremum of
the density, is $\left|\nabla \rho\right|^m$, then
\begin{equation}
\frac{\partial\, e_{\rm xc}}{\partial\,\nabla \rho} \sim
\left|\nabla \rho\right|^{m-1} \sim r^{(s-1)(m-1)},
\end{equation}
and
\begin{equation}
\nabla\cdot\left(\frac{\partial\, e_{\rm xc}}{\partial\,\nabla \rho} \right)
\sim r^{(s-1)(m-1)-1}.
\end{equation}
Therefore, the exchange-correlation potential diverges at the origin if
\begin{eqnarray}
(s-1)(m-1)-1<0, \label{condition}
\end{eqnarray}
where {\it s} and {\it m} are positive.
Thus, the exchange-correlation potential diverges at extrema of the density
for all values of {\it s} if $m\leq 1$, as is the case for the Wilson-Levy
potential~\cite{WL}, and for all values of {\it m} if $s\leq 1$.
At nuclei, $s=1$ so that the exchange-correlation potential always diverges.

As far as the large distance asymptotics are concerned, any GGA that includes
no higher than first derivatives of the density cannot simultaneously satisfy
both the correct $-1/r$ behavior of the exchange-correlation potential and the
correct $-1/2r$ behavior of the local exchange-correlation energy per
electron~\cite{ECMV}.

It is however possible, as mentioned in Refs.~\cite{proc,he}, to include the
Laplacian of the density in an appropriate way~\cite{U_notes} and construct a GGA that has
the triple advantage that it satisfies both of the long-distance asymptotic
conditions and also does not suffer from a spurious divergence at the nuclei.
There are an infinite number of possible variations in the functional form
that preserve the desired short and long range asymptotic behaviors.
We are presently searching for the most physically reasonable functional
expression and for a universal set of parameter values in such a functional.
In the same line of research, \"Jemmer and Knowles have recently published a
simple expression for an exchange functional depending on the Laplacian of
the density and reproducing the correct long and short distance asymptotics
of the exchange-correlation potential and local exchange-correlation
energy per electron~\cite{JK}.
However, they conclude that such a functional is not a suitable choice
as a general purpose density functional.
The importance of Laplacian terms has also been pointed out by Engel and
Vosko~\cite{EV2} who show that inclusion of Laplacian terms in a fourth order
Taylor expansion of the exchange-correlation functional results in improved
exchange potentials for atoms and jellium spheres.
However, their expression does not obey any of the above three asymptotic
conditions.

\section{DETERMINATION OF ACCURATE DFT QUANTITIES}
\label{s3}

For atomic systems, it is possible to determine accurate exchange-only
quantities by using the solution of the optimized effective potential (OEP)
method which represents the exact solution in an exchange-only DFT
scheme~\cite{OEP}. However, these quantities are determined not for the exact
density but for the OEP density which corresponds to the self-consistent
solution in an exchange-only approach.
The performance of exchange-only approximate functionals and the corresponding
exchange potentials can therefore be routinely checked for atomic systems by
using the OEP method. Approximate quantities in exchange-only GGA, evaluated
either for the OEP density or the self-consistent density, can be compared
with the corresponding OEP quantities.

Obtaining reliable exchange-correlation potentials and energies is instead a
more difficult task.  One has to generate an accurate density and then
compute an exchange-correlation potential that yields the desired density
as a solution of the Kohn-Sham equations (Eq.~\ref{KS} and~\ref{rho}).
In this context, researchers have used charge densities, of varying
degrees of accuracy, calculated by quantum chemistry methods for
atoms~\cite{S79,vonBarth84,AP,AS,NM,D90,Chen94,LB1,MZ,Ludena95} and
molecules~\cite{BBS,GLB,IH}, as well as Quantum Monte Carlo methods for
atoms~\cite{proc,UG} and for a model semiconductor~\cite{KG}.
The subsequent inverse problem, namely the search of the corresponding
exchange-correlation potential, has been performed using a variety of different
techniques. For example, in the special case of the singlet ground state of a
two-electron system, the exchange-correlation potential can be obtained
simply from Eq.~\ref{KS} while, for systems with more than two electrons,
$v_{\rm xc}$ can be determined by expanding it in a complete set of basis
functions and varying the expansion coefficients such that Eqs.~\ref{KS}
and \ref{rho} yield the accurate density\cite{AP,UG}.
Whatever the technique used, an accurate density is a key ingredient
for the determination of the exchange-correlation potential, because small
errors in the density are greatly magnified by the inversion procedures used
to obtain the potential.

By knowing the exchange-correlation potential, we are able to calculate the
Kohn-Sham orbitals and, if the total energy of the system can be estimated,
the exchange-correlation energy is obtained by inversion of the expression
for the total energy (Eq. \ref{eq0}):
\begin{eqnarray}
E_{\rm xc}\left[\rho\right]=E
-\frac{1}{2}\sum_{i=1}^N\int\psi_i\nabla^2\psi_i\,{\rm d}{\bf r}
-\frac{1}{2}\int\!\!\int \frac{\rho({\bf r})
\rho({\bf r}')}{\left|{\bf r}-{\bf r}'\right|}{\rm d}{\bf r}\,{\rm d}{\bf r}'
-\int\,\rho\left({\bf r}\right)v_{\rm ext}({\bf r}){\rm d}{\bf r}.
\label{excf}
\end{eqnarray}
Since the Kohn-sham orbitals are known, the exchange energy (Eq.~\ref{enx0})
can be calculated as
\begin{eqnarray}
E_{\rm x}\left[\rho\right]=
-\frac{1}{2}\sum_{i=1}^N\sum_{j=1}^N\delta_{m_{s_i},m_{s_j}}\int\!\!\int
\frac{\psi_i^*({\bf r})\psi_j^*({\bf r}')\psi_j({\bf r})\psi_i({\bf r}')}
{\left|{\bf r}-{\bf r}'\right|}\,{\rm d}{\bf r}\,{\rm d}{\bf r}',
\label{enx1}
\end{eqnarray}
where the $\delta$-function is over the spin quantum numbers of the $i$-th
and $j$-th spin-orbitals. The correlation energy is finally obtained as
the difference of the exchange-correlation energy (Eq.~\ref{excf}) and the
exchange energy (Eq.~\ref{enx1}):
\begin{eqnarray}
E_{\rm c}\left[\rho\right]=E_{\rm xc}\left[\rho\right]-
E_{\rm x}\left[\rho\right].
\end{eqnarray}

The functional derivative of the exchange energy (Eq.~\ref{enx1}) with
respect to the density yields the exchange potential.
By knowing the exchange-correlation and the exchange potentials, we obtain the
correlation potential as the difference
\begin{eqnarray}
v_{\rm c}\left(\left[\rho\right];{\bf r}\right)=
v_{\rm xc}\left(\left[\rho\right];{\bf r}\right)-
v_{\rm x}\left(\left[\rho\right];{\bf r}\right).\label{vc}
\end{eqnarray}
We show in Sub-sec.~\ref{s3.1} that it is easy to calculate the exchange
potential for a system of two electron in a singlet state. The procedure
is instead much more involved for a many-electron system, as we outline in
Sub-sec.~\ref{s3.2}.

As already introduced in Sec.~\ref{s0}, the exchange-correlation functional
can be written as an integral over the exchange-correlation energy
density, $e_{\rm xc}\left({\bf r}\right)$, or the local exchange-correlation
energy per electron, $\epsilon_{\rm xc}\left({\bf r}\right)$:
\begin{eqnarray}
E_{\rm xc}\left[\rho\right]
=\int e_{\rm xc}\left({\bf r}\right) {\rm d}{\bf r}
=\int \rho\left({\bf r}\right) \epsilon_{\rm xc}\left({\bf r}\right) {\rm d}{\bf r}.
\end{eqnarray}
The separation of $E_{\rm xc}$ into exchange and correlation
(Eq.~\ref{separaE}) corresponds to an equivalent decomposition of
$\epsilon_{\rm xc}$ into $\epsilon_{\rm x}$ and $\epsilon_{\rm c}$.
The definition of $\epsilon_{\rm xc}$ and its components is however not unique
since two $\epsilon_{\rm xc}$'s whose difference is a function $f$ with
$\int \rho\left({\bf r}\right) f\left({\bf r}\right){\rm d}{\bf r}=0$
yield the same exchange-correlation energy functional.

We observe that a natural definition of the local exchange energy per electron
follows from Eq.~\ref{enx1}:
\begin{eqnarray}
\epsilon_{\rm x}({\bf r})=-\frac{1}{2}
\frac{1}{\rho({\bf r})}\sum_{i=1}^N\sum_{j=1}^N\delta_{m_{s_i},m_{s_j}}
\psi_i^*({\bf r})\psi_j({\bf r})\int \frac{\psi_j^*({\bf r}')\psi_i({\bf r}')}
{\left|{\bf r}-{\bf r}'\right|}\,{\rm d}{\bf r}'.
\label{exden}
\end{eqnarray}
This definition results in an $\epsilon_{\rm x}$ that at large distances goes
as $-1/2r$ \cite{M}. A procedure for constructing $\epsilon_{\rm c}$ has been
recently proposed using first and second order density matrices from correlated
wave functions and applied to He and H$_2$ \cite{SGNB}.

In the remaining sections, the accurate exchange-correlation energies and
potentials, the separate exchange and correlation components and the local exchange
energy per electron are used to test the accuracy of the approximate
exchange-correlation energy functionals listed in Appendix~\ref{a1}.

\subsection{Two-electron systems}
\label{s3.1}

For two electrons of opposite spin, there is a single spatial Kohn-Sham orbital
$\psi$ and is simply related to the electronic density (Eq.~\ref{rho}) as
\begin{eqnarray}
\psi({\bf r})=\left[\frac{\rho({\bf r})}{2}\right]^{1/2}.
\end{eqnarray}
Knowing the exact density, the exchange-correlation potential can be
obtained by inversion of the Kohn-Sham equation (Eq.~\ref{KS}),
\begin{eqnarray}
v_{\rm xc}(\left[\rho\right];{\bf r})=\epsilon_{\rm KS}
       +\frac{1}{2}\frac{\nabla^2\psi}
{\psi}-v_{\rm ext}({\bf r})-\int \frac{\rho({\bf r}')}{\left|{\bf r}-{\bf r}'
\right|} {\rm d}{\bf r}'.
\end{eqnarray}
The eigenvalue $\epsilon_{\rm KS}$ is equal to minus the ionization energy
if we impose that the exchange-correlation potential goes to zero at infinity.
Therefore, the exchange-correlation potential is completely determined.
From Eq.~\ref{enx1}, it follows that the exchange energy is given by
\begin{eqnarray}
E_{\rm x}\left[\rho\right]=-\frac{1}{4}\int\int
\frac{\rho({\bf r}) \rho({\bf r}')}{\left|{\bf r}-{\bf r}'\right|}
{\rm d}{\bf r} \,{\rm d}{\bf r}',\label{natural}
\end{eqnarray}
and the functional derivative of this expression with respect to the density
yields the exact exchange potential:
\begin{eqnarray}
v_{\rm x}\left(\left[\rho\right];{\bf r}\right)=
-\frac{1}{2}\int\frac{\rho({\bf r}')}
{\left|{\bf r}-{\bf r}'\right|} {\rm d}{\bf r}'.\label{vx}
\end{eqnarray}
Observe that $v_{\rm x}\left(\left[\rho\right];{\bf r}\right)$ simply
follows from the condition that it cancels the self-interaction term in the
Hartree potential.

For a system of two interacting electrons in a singlet state, Eq.~\ref{exden}
for the local exchange energy per electron reduces to
\begin{eqnarray}
\epsilon_{\rm x}({\rm r})=-\frac{1}{4}\int\frac{\rho\left({\bf r}'\right)}
{\left|{\bf r}-{\bf r}'\right|}{\rm d}{\bf r}'.
\end{eqnarray}
The same expression could have also been obtained from Eq.~\ref{natural}.

\subsection{Many-electron systems}
\label{s3.2}

While the separation of the exchange-correlation potential into exchange plus
correlation is quite simple in the two-electron systems, this decomposition
according to the DFT definition has never been obtained for many-electron
systems.
In previous work~\cite{AP,AS,Chen94,proc,Ludena95}, the
exchange potential was defined as the difference of the effective Kohn-Sham
potential yielding the Hartree-Fock density and the sum of the Hartree
and the external potentials.
The correlation potential was then obtained as the difference of the
exchange-correlation potential corresponding to the exact density and the
above potential.
Note that this separation does not correspond to the DFT definition of
Eq.~\ref{vc}: it involves two densities,
the exact and the Hartre-Fock densities, while the potential used
for exchange is only approximately equal to the exchange potential
corresponding to the Hartree-Fock density
(although very close to it) since it is not the functional derivative with
respect to the density of the exchange energy evaluated for the orbitals
obtained from the effective potential yielding the Hartree-Fock density.
Here, we obtain the correct separation of the exchange and correlation
components, according to their DFT definition, for the exact density.
The difference between the exchange potential we obtain and the approximate
``exchange'' potential described above is almost not detectable on the scale
of the exchange potential. On the other hand, on the more expanded scale of
the correlation potential, the difference is clearly visible,
although the shapes are very similar.

We follow G\"orling and Levy \cite{GL} in showing how to
separate the exchange-correlation potential into exchange and correlation.
We consider a spin unpolarized system.
If we assume that the density $\rho$ is non-interacting {\it v}-representable,
it can be expressed as in Eq.~\ref{rho} in terms of single-particle orbitals
$\{\psi_i\}$ of the Kohn-Sham potential $v_{\rm s}\left({\bf r}\right)$,
\begin{eqnarray}
v_{\rm s}\left({\bf r}\right)=v_{\rm ext}\left({\bf r}\right)+
\int\frac{\rho\left({\bf r}'\right)}{\left|{\bf r}-{\bf r}'\right|}
{\rm d}{\bf r}'+v_{\rm xc}\left(\left[\rho\right];{\bf r}\right).
\end{eqnarray}
We evaluate the functional derivative of the exchange energy functional
with respect to the Kohn-Sham potential as
\begin{eqnarray}
\frac{\delta E_{\rm x}\left[\rho\right]}{\delta v_{\rm s}({\bf r})}=
\int\frac{\delta E_{\rm x}\left[\rho\right]}{\delta
\rho({\bf r}')}\frac{\delta \rho({\bf r}')}{\delta v_{\rm s}({\bf r})}\,
{\rm d}{\bf r}'
=\int v_{\rm x}\left(\left[\rho\right];{\bf r}'\right)
\sum_{i=1}^N\left(\psi_i^*({\bf r}')\frac{\delta \psi_i({\bf r}')}
{\delta v_{\rm s}({\bf r})}+\frac{\delta \psi_i^*({\bf r}')}
{\delta v_{\rm s}({\bf r})}\psi_i({\bf r}')\right)\,
{\rm d}{\bf r}'.
\label{oep1}
\end{eqnarray}
On the other hand, since the exchange functional can be written as a function
of the orbitals (Eq.~\ref{enx1}), we also have
\begin{eqnarray}
\frac{\delta E_{\rm x}\left[\rho\right]}{\delta v_{\rm s}({\bf r})}=
\sum_{i=1}^N\int\left(\frac{\delta E_{\rm x}\left[\rho\right]}
{\delta \psi_i({\bf r}')}\frac{\delta \psi_i({\bf r}')}
{\delta v_{\rm s}({\bf r})}
+\frac{\delta E_{\rm x}\left[\rho\right]}{\delta \psi_i^*({\bf r}')}
\frac{\delta \psi_i^*({\bf r}')}{\delta v_{\rm s}({\bf r})}\right)\,
{\rm d}{\bf r}'.
\label{oep2}
\end{eqnarray}
If we combine Eqs.~\ref{oep1} and \ref{oep2}, we obtain the integral equation
\begin{eqnarray}
\int v_{\rm x}\left(\left[\rho\right];{\bf r}'\right)
{\cal K}\left({\bf r}',{\bf r}\right){\rm d}{\bf r}'=
{\cal Q}\left({\bf r}\right),
\label{vxeq}
\end{eqnarray}
where the kernel ${\cal K}\left({\bf r}',{\bf r}\right)$ and the right hand
side ${\cal Q}\left({\bf r}\right)$ depend on the orbital $\{\psi_i\}$ and
their functional derivative with respect to the potential $v_{\rm s}\left(
{\bf r}\right)$.
This integral equation is equivalent to the one solved in the OEP method with
the KS orbitals replaced by the OEP orbitals~\cite{OEP}.
The functional derivatives of the orbitals $\delta \psi_i({\bf r})/
\delta v_{\rm s}({\bf r}')$ can be expressed in terms of the Green's function
$G_i\left({\bf r},{\bf r}'\right)$ as
\begin{eqnarray}
\frac{\delta \psi_i({\bf r})}{\delta v_{\rm s}({\bf r}')}
=-G_i\left({\bf r},{\bf r}'\right)\psi_i({\bf r}'),
\end{eqnarray}
where $G_i\left({\bf r},{\bf r}'\right)$ satisfies the differential equation
\begin{eqnarray}
\left(-\frac{1}{2}\nabla^2+v_{\rm s}({\bf r})-\epsilon_i\right)
G_i\left({\bf r},{\bf r}'\right)
=\delta\left({\bf r}-{\bf r}'\right)-\psi_i({\bf r})\psi_i^*({\bf r}').
\end{eqnarray}
By knowing the exchange-correlation potential, the KS orbitals and eigenvalues,
we can compute the Green's functions $\{G_i\}$ and, consequently, the kernel
$\cal K$ and the function $\cal Q$.
If we express the exchange potential as a linear combination of basis
functions, Eq.~\ref{vxeq} can be rewritten as a non-homogeneous set of
linear equations for the coefficients of the expansion of the potential
in the basis set. More details on the separation procedure can be found
in Ref.~\cite{CCX}.

\section{SYSTEMS STUDIED AND CORRESPONDING DENSITIES}
\label{s4}

Experimental and computed energies are often compared in order to assess the
accuracy of approximate density functionals. However, this procedure is not
very reliable since energies are integrated quantities and are therefore
subject to cancellation of errors.
In order to better understand the performance of approximate
exchange-correlation functionals, we will examine not only total energies
but also the components of the energy, the self-consistent charge densities,
the exchange-correlation potentials and the local exchange energies per electron.
We will compare these quantities, obtained from approximate schemes, with
the corresponding accurate quantities derived from (1) the exact solution of a
model system of two interacting electrons in a harmonic potential (we call
this system ``Harmonium''), (2) a nearly exact wave function for the He atom
and (3) accurate quantum Monte Carlo calculations for the Be and Ne atoms.

\subsection{Harmonium}
\label{s4.1}

\begin{figure}[htbp]
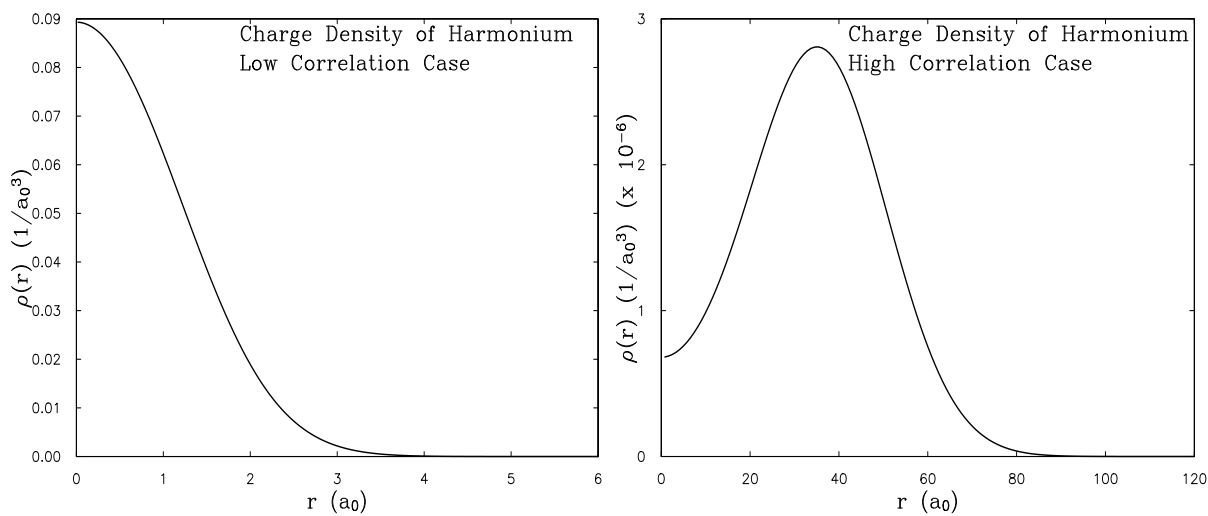

\centerline{{\epsfxsize=7.9 cm \epsfysize=6.7 cm \epsfbox{pl.harm2.rho}}
            {\epsfxsize=7.9 cm \epsfysize=6.7 cm \epsfbox{pl.harm10.rho}}}
\caption[]{Ground state electronic densities: the spring constants are
$k=0.25$ a.u. (left, low correlation case) and $k\approx 3.6\times 10^{-6}$
a.u. (right, high correlation case). Taken from Ref.~\cite{harmonium} and
modified.}
\label{harmd10}
\end{figure}
A simple two-electron system where the exchange-correlation potential and
energy is calculable exactly consists of two interacting electrons in a
harmonic potential. The model can be solved analytically for a discrete and
infinite set of values of the spring constant $k$~\cite{T} and the degree of
correlation within the system can be varied by simply tuning the value of $k$.

In Fig. \ref{harmd10}, we show the density in a low correlation case,
corresponding to a spring constant $k=0.25$ a.u., and in a high correlation
case with a spring constant $k\approx 3.6\times 10^{-6}$ a.u.
In the low correlation case, the density has a maximum at the origin while,
in the high correlation case, the density has a local minimum at the origin
and an absolute maximum at a finite distance from the origin~\cite{harmonium}.
Although the high correlation density is too low to be physically relevant for
electronic structure calculations ($\left<r_s\right>=57.9$ a.u.), the system
is interesting because it is strongly correlated and its density differs
qualitatively from an atomic density due to the presence of a maximum
at a finite distance from the origin. Consequently, in the present paper, we
will only discuss the high correlation limit. For an analysis of the system
with $k=0.25$ a.u., see Ref.~\cite{harmonium}.

\subsection{He atom}
\label{s4.2}

\begin{figure}[htbp]
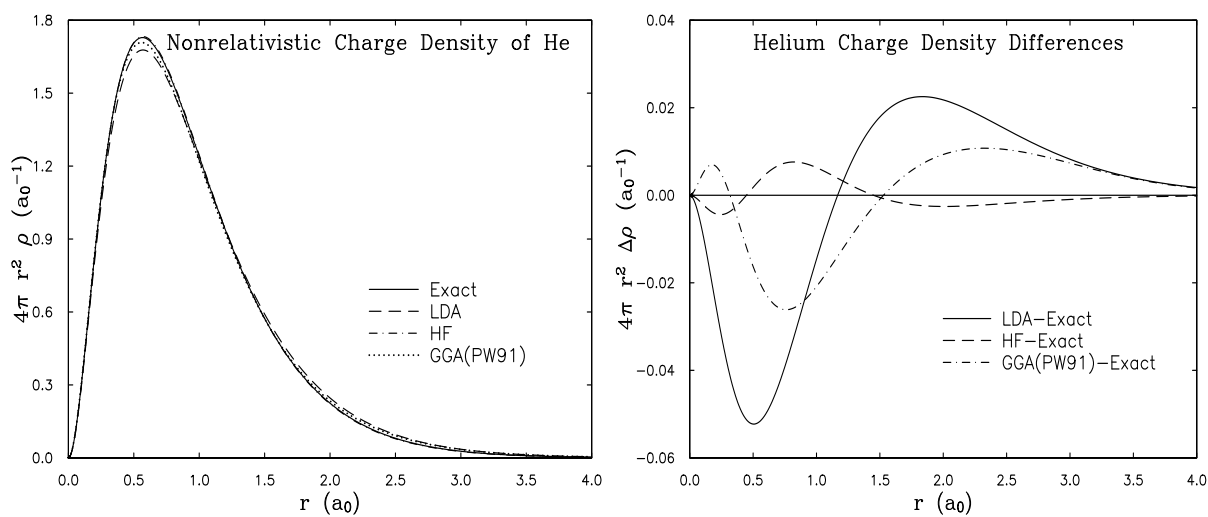

\centerline{{\epsfxsize=7.9 cm \epsfysize=6.7 cm \epsfbox{pl.he.rho}}
            {\epsfxsize=7.9 cm \epsfysize=6.7 cm \epsfbox{pl.he.delrho}}}
\caption[]{Comparison of the LDA, GGA (PW91), HF and exact densities of He
(left). Errors in the LDA, GGA, and HF densities of He (right). Taken from
Ref.~\cite{he} and modified.}
\label{he.density}
\end{figure}
It is necessary to employ very accurate wave functions in order to observe the
true short-distance and long-distance behavior to ${\cal O} (1/r)$ of the
exchange-correlation potential. For the He iso-electronic series, it is
possible to achieve this accuracy. The form of the wave function is
described in Refs.~\cite{he,Morgan}. The energy for He corresponding to
this wave function was estimated to be about 1 part in 10$^{16}$.
With this nearly exact wave function, the exchange-correlation potential
is resolved with unprecedenced accuracy: it is possible to observe~\cite{he}
not only the $-1/r$ behavior of the exchange potential but also the
asymptotic behavior of the correlation potential, $-9/(4 Z^4 r^4)$,
predicted by Almbladh and von Barth~\cite{AB}.

In Fig.~\ref{he.density}, we compare the density obtained with our
accurate wave function with the self-consistent densities obtained from
LDA, the Perdew-Wang '91 GGA\cite{PW91} and Hartree-Fock (HF). The LDA density
is less peaked than the true density while the HF density is very close but
slightly more peaked than the true density.
We also show the error in the LDA, GGA, and HF densities.
The GGA density is more accurate than the LDA density but less accurate than
the HF one.
Note that the error in the HF density is considerably smaller than the LDA and
GGA errors; however, as shown below, this is true only in the core region for
heavier systems.

\subsection{Be atom}
\label{s4.3}

For many-electron systems, an accurate determination of the charge densities
is obtained by combining the results from variational Monte Carlo (VMC) and
diffusion Monte Carlo (DMC) methods~\cite{UNR}. In VMC, the square of a trial
wave function is sampled and the expectation values for the trial wave function
are calculated as averages over the sampled configurations. An advantage of
this technique versus other conventional quantum chemistry methods is that the
many-dimensional integrals can be performed for any given form of the trial
wave function. Therefore, one is freed from the constraint of having to
express the trial wave function in some restricted form such as a linear
combination of determinants of single-particle orbitals.
The starting point of the DMC method is also a good trial wave function from
which DMC projects out an improved estimate of the true wave function.

To calculate the charge density, the integral of the square of the true wave
function over all but one of the electrons has to be estimated. VMC samples
the square of the trial wave function. DMC samples the product of the trial
wave function and the true wave function.
If the error in the wave function is of ${\cal O}(\epsilon)$, the errors in
the densities obtained from VMC and DMC are of the same order.
A density with an error of ${\cal O}(\epsilon^2)$ (that we will refer to as
the quantum Monte Carlo (QMC) density) is obtained by taking twice the DMC
density and subtracting the VMC density\cite{gfmcbook}.
The QMC density is obtained in the form of an histogram and then fitted to a
sum of products of monomials, exponentials and an appropriate asymptotic 
function~\cite{Patil}.

\begin{figure}[htbp]
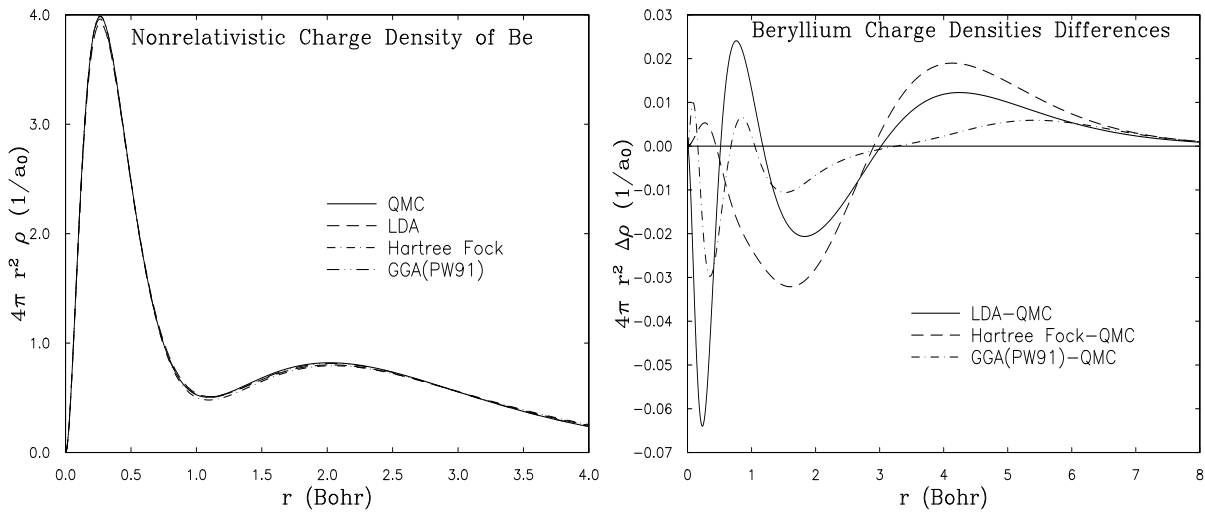

\centerline{{\epsfxsize=7.9 cm \epsfysize=6.7 cm \epsfbox{pl.be.rho}}
            {\epsfxsize=7.9 cm \epsfysize=6.7 cm \epsfbox{pl.be.delrho}}}
\caption{Comparison of the LDA, GGA (PW91), HF and ``exact" (QMC) densities of Be
(left). Errors in the LDA, GGA, and HF densities of Be (right). Taken from
Ref.~\cite{UG}.}
\label{be.den}
\end{figure}

In Fig.~\ref{be.den}, we show the densities (left) obtained from LDA, the
Perdew-Wang '91 GGA, HF and QMC for Be and the errors in these
densities (right), with the QMC density as the reference.
The first thing to notice is that all densities are very similar.
The LDA and HF errors are of comparable magnitude in the valence region
whereas, as in the case of He (Fig.~\ref{he.density}), the HF error is smaller
than the LDA error in the region near the nucleus.
A hand-waving explanation of this behavior is that, in the core region,
exchange dominates correlation and HF, by definition, is exact for exchange
only.
The Perdew-Wang '91 density is somewhat more accurate than the LDA density.
The GGA potential has a spurious negative divergence at the nucleus that is
partially responsible for the improvement in the density.
The LDA density is too low at the nucleus and the negative divergence of the
GGA potential results in increased charge density there, a feature we have
observed for Ne as well.

\section{A CLOSER LOOK AT EXISTING APPROXIMATIONS: {\Large $v_{\rm xc}$},
{\Large $v_{\rm x}$} AND {\Large $v_{\rm c}$}}
\label{s5}

The accurate exchange-correlation potentials and their separation in
exchange and correlation components were obtained for the four systems
described in this paper, using the techniques mentioned in
Sec.~\ref{s3}.
The approximate exchange-correlation potentials are evaluated numerically as
functional derivatives of the approximate exchange-correlation energy
functionals (Eq.~\ref{pot}) and compared with the corresponding
accurate quantities.

\subsection{Harmonium}
\label{s5.1}

\begin{figure}[htbp]
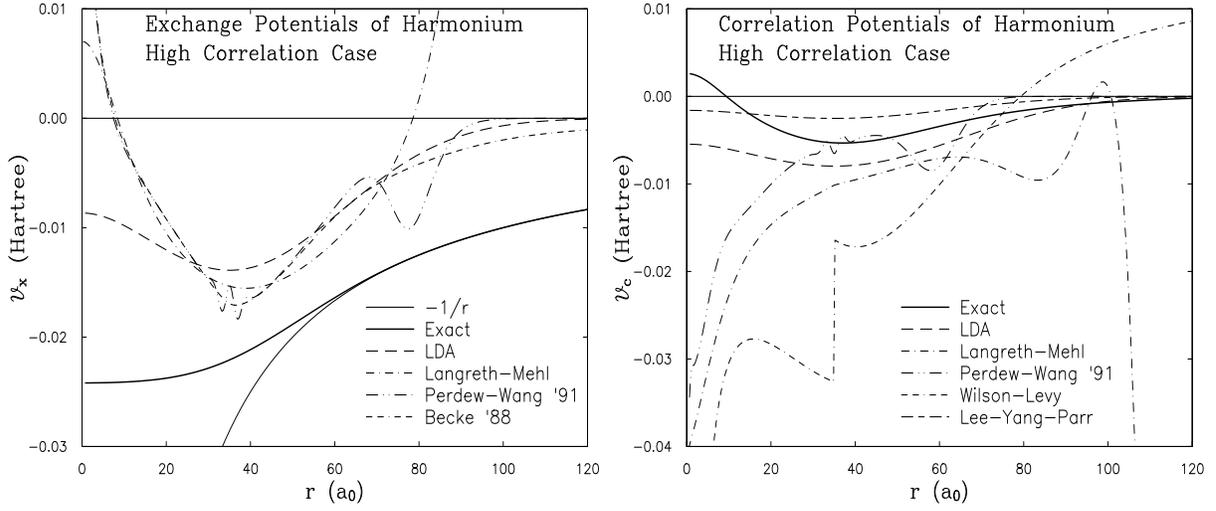

\centerline{{\epsfxsize=7.9 cm \epsfysize=6.7 cm \epsfbox{pl.harm10.vx}}
            {\epsfxsize=7.9 cm \epsfysize=6.7 cm \epsfbox{pl.harm10.vc}}}
\caption[]{Exact and approximate exchange (left) and correlation (right)
potentials for the spring constant $k\approx 3.6\times 10^{-6}$ a.u.
Taken from Ref.~\cite{harmonium} and modified.}
\label{vxc10}
\end{figure}

In Fig. \ref{vxc10}, we plot the exact exchange potential and several
approximate exchange potentials evaluated for the exact density of the model
system.
The approximate potentials differ from the exact potential over all the spatial
range. At large {\it r}, all the potentials do not reproduce the correct $-1/r$
asymptotic behavior: they go exponentially to zero with the exception of the
Becke functional which behaves as $(-{\rm constant}/r^2)$~\cite{ECMV} and of
the Langreth-Mehl functional which diverges.
At the origin, the exact exchange potential has a minimum because it is
simply proportional to the Hartree potential through Eq.~\ref{vx}
while the approximate exchange potentials exhibit a maximum (they do not
diverge since, from Eq.~\ref{condition}, $s=2$ and $m>1$).
The Perdew-Wang '91 exchange potential is very close to the Becke
'88 potential except that it displays additional oscillations
at the extrema of the densities and in the tail of the density.
In addition, as mentioned earlier, the long-distance asymptotic behaviors
are different.
As pointed out by Engel and Vosko~\cite{EV1}, the oscillations of the
Perdew-Wang '91 potential at the quadratic extrema of the density are due to
the fact that the effective coefficient of the $\vert\nabla\rho\vert^2$ term
increases very rapidly from the known value at $\xi=0$, specified by the
second-order gradient expansion, to a 2.1 times larger value at only
$\xi=0.04$, where $\xi=[\nabla\rho/(2 k_F \rho)]^2$ and $k_F$ is the Fermi
wave vector. Observe that the Becke '88 potential does not have spurious
oscillations at the extremum of the density nor in the tail of density.

In Fig. \ref{vxc10}, we also show the exact correlation potential and several
approximate correlation potentials evaluated for the exact density.
At large {\it r}, the approximate potentials go to zero exponentially with the
exception of the Wilson-Levy potential which goes to a positive
constant and the Langreth-Mehl functional which diverges.
As predicted by Eq.~\ref{condition}, the Wilson-Levy potential
diverges at the origin: the energy density,
$e_{\rm xc}(\rho,\left|\nabla\rho\right|)$, depends on $\left|\nabla\,
\rho\right|$ as ${\cal O}(\left|\nabla \rho\right|))$ for small values of the
gradient of the density (Eq.~\ref{condition}, $m=1$). All the other functionals
assume a finite value at the origin (Eq.~\ref{condition}, $s=2$ and $m>1$).
The Wilson-Levy functional has a discontinuity at the maximum of the electronic
density: the functional derivative of the exchange-correlation energy contains
the sign of the radial derivative of the density and is discontinuous if the
energy density depends on the absolute value of gradient of the density as
${\cal O} (\left|\nabla \rho\right|))$, for small values of
$\left|\nabla \rho\right|$.  As in the case of exchange
the Perdew-Wang '91 correlation potential shows an oscillatory behavior near
the maximum at a finite distance from the origin and also in the tail of
the potential.


\subsection{He atom}
\label{s5.2}

For atomic systems, the exchange potential is the dominant part of the
exchange-correlation potential. It shares with it most of its characteristics:
the asymptotic behavior, the intershell bump (see Sub-sec.~\ref{s5.3}) and the
finite short-range behavior.

\begin{figure}[htbp]
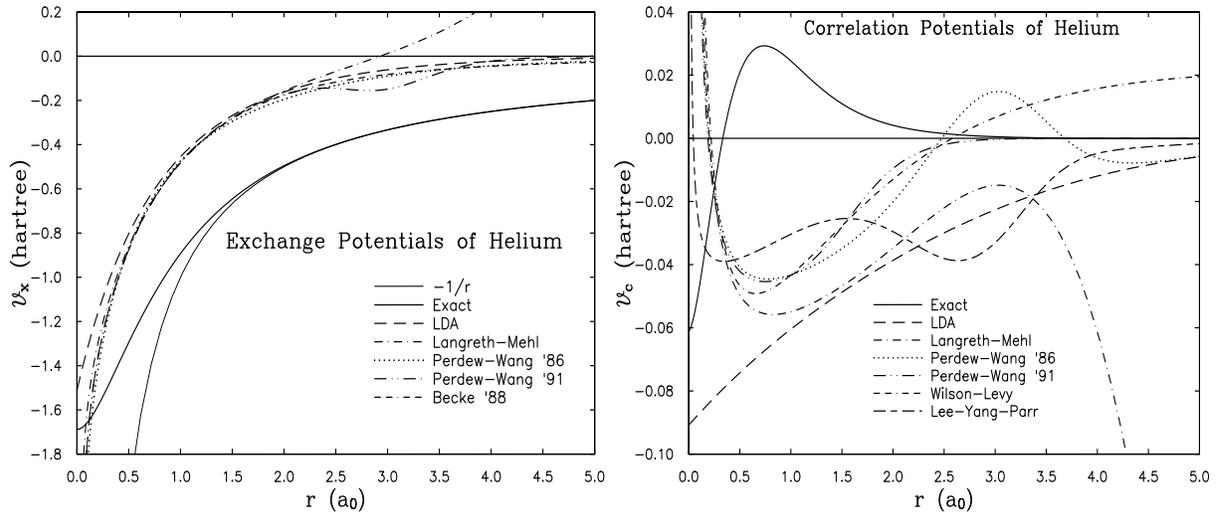

\centerline{{\epsfxsize=7.9 cm \epsfysize=6.7 cm \epsfbox{pl.he.vx}}
            {\epsfxsize=7.9 cm \epsfysize=6.7 cm \epsfbox{pl.he.vc}}}
\caption[]{Exact and approximate exchange (left) and correlation (right)
potentials for He. Taken from Ref.~\cite{he} and modified.}
\label{vxc.he}
\end{figure}

In Fig.~\ref{vxc.he} we compare the exchange potential obtained from the LDA
and the various GGA's evaluated for the exact charge density with the exact
potential obtained from Eq.~\ref{vx}.
The approximate potentials differ significantly from the exact one.
At large $r$, all the potentials do not reproduce the correct $-1/r$ asymptotic
behavior: they go exponentially to zero with the exception of the Becke '88
functional which behaves as $(-{\rm constant}/r^2)$~\cite{ECMV}
and of the Langreth-Mehl functional which diverges.
The Perdew-Wang '91 functional has a spurious minimum between 2.5 and 3.5
$a_0$.

At the origin, the exact exchange potential has a quadratic minimum because it
is simply proportional to minus the Hartree potential. In contrast all the
proposed GGA exchange potentials diverge at the nucleus while the LDA exchange
potential has a finite value and slope there.
Hence, very close to the nucleus, the various GGA potentials are an even
poorer approximation to the true potential than LDA but the more negative
values of the GGA potentials at short and intermediate distances is a step
in the right direction. As shown in Eq.~\ref{condition}, any approximate
functional that has terms containing the gradient but no higher derivatives
of $\rho$ must yield an exchange-correlation potential that diverges at nuclei
since, at nuclei, $s=1$.

In Fig.~\ref{vxc.he}, we also compare approximate correlation potentials of He
to the exact one. In the case of an $n$-electron atom or ion with orbitally
non-degenerate $n$ and $n-1$ electron ground states, the correlation potential
goes as $-\alpha/2r^4$ at large distances, where $\alpha$ is the
dipole-polarizability of the $n-1$ electron system\cite{AB}.
In the case of the He iso-electronic series, the dipole-polarizability of the
residual 1-electron system is $\alpha=9/(2 Z^4)$.
Due to the rapid fall-off of $1/r^4$, this
asymptotic behavior is not discernible in Fig.~\ref{vxc.he} but is
evident in a plot of $r^4v_{\rm c}$~\cite{he}.
At large distances, the approximate potentials go to zero exponentially with
the exception of the Wilson-Levy potential which goes to a positive
constant and the Langreth-Mehl potential which diverges.
All the GGA correlation potentials diverge at the origin as follows
from Eq.~\ref{condition}.
Of the GGA's proposed so far, only the Lee-Yang-Parr GGA contains Laplacian
terms but not in the form necessary to eliminate the divergence at the
nucleus.

\subsection{Be and Ne atoms}
\label{s5.3}

\begin{figure}[htbp]
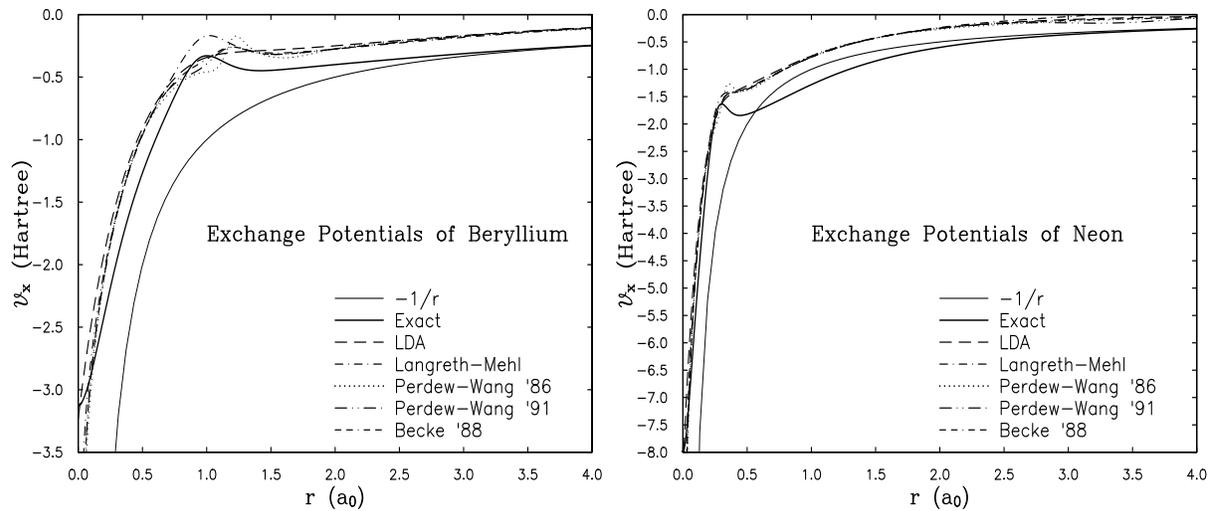

\centerline{{\epsfxsize=7.9 cm \epsfysize=6.7 cm \epsfbox{pl.be.vx}}
            {\epsfxsize=7.9 cm \epsfysize=6.7 cm \epsfbox{pl.ne.vx}}}
\caption{Exchange potentials of Be (left) and Ne (right) from QMC
(labeled exact - see text), LDA and various GGA's. Taken from Ref.~\cite{UG}.}
\label{BeNe.vx}
\end{figure}

In Fig. \ref{BeNe.vx}, we show the accurate 
exchange potentials of Be and Ne and compare them
with the approximate exchange potentials from the local density approximation
and several generalized gradient approximations, all of them being evaluated
for the corresponding accurate QMC density.

\begin{figure}[htbp]
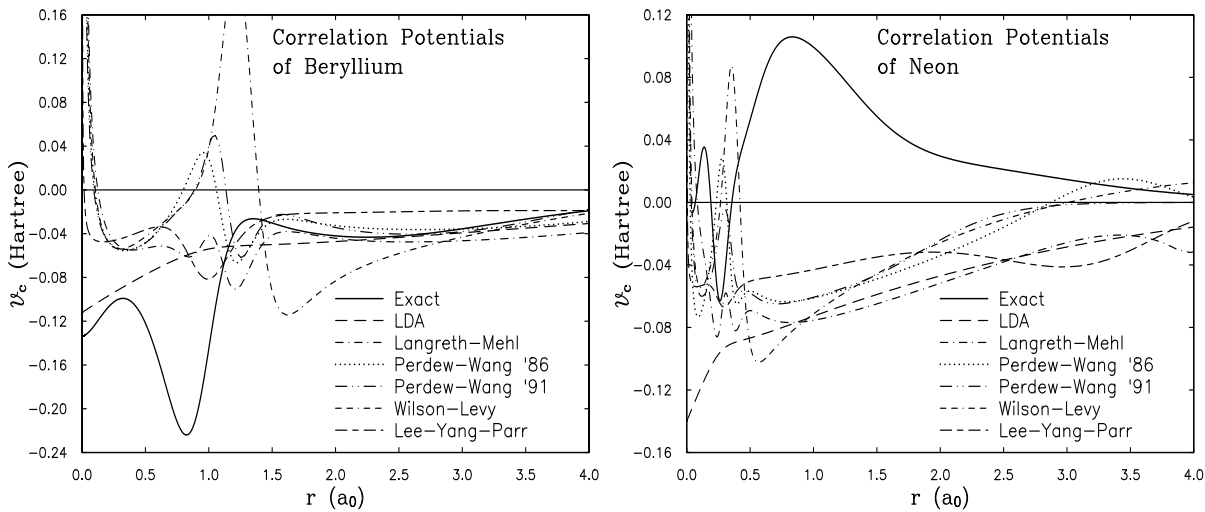

\centerline{{\epsfxsize=7.9 cm \epsfysize=6.7 cm \epsfbox{pl.be.vc}}
            {\epsfxsize=7.9 cm \epsfysize=6.7 cm \epsfbox{pl.ne.vc}}}
\caption{Correlation potentials of Be (left) and Ne (right) from QMC
(labeled exact - see text), LDA and various GGA's. Taken from Ref.~\cite{UG}.}
\label{BeNe.vc}
\end{figure}

At short and large distances, all GGA's are similar and qualitatively wrong
and the same analysis applies as in the case of the He atom.
The Perdew-Wang '91 exchange potential
has a spurious dip at about 6 a.u. (outside the plotted range) for Be and
at about 3.3 a.u. for Ne. This dip is not present in either Perdew-Wang '86
or the Becke '88 potentials.
The exact exchange potential has a peak in the intershell region (around 1 a.u.
for Be and around 0.3 a.u. for Ne).
There, the various GGA potentials exhibit some differences but they all have
a peak at approximately the right position.
The GGA potentials are more accurate than the LDA potential at distances
smaller than the intershell radius, except at very short distances where they
have a spurious divergence.
To summarize, the exchange potentials from the GGA's show some tendency for
improvement upon LDA in the intershell region and at shorter distances but
they have a spurious divergence at the nucleus and do not improve the
long-distance behavior.

In Fig.~\ref{BeNe.vc}, we show the correlation potentials for Be and Ne from QMC,
LDA and the different GGA's.
The LDA correlation potential is everywhere negative, monotonically
increasing with the radius (since the density is monotonically decreasing) and
rather smooth. Except for the part of the potential beyond 1.5 a.u. for Be, it
bears no resemblance to the exact correlation potential. The GGA correlation
potentials exhibit more structure but none of these does better than LDA.
Nevertheless, as we will see in Section~\ref{s7} the GGA's yield more accurate
correlation energies than does LDA.
The true correlation potential is everywhere negative for Be but
is predominantly positive for Ne.

\section{LOCAL EXCHANGE ENERGY PER ELECTRON: {\Large $\epsilon_{\rm x}$}}
\label{s6}

Using the expression in Eq.~\ref{exden} for the local exchange energy per
electron, $\epsilon_{\rm x}$, we present accurate $\epsilon_{\rm x}$'s
for the two-electron systems and the Be atom.
Since $\epsilon_{\rm x}$ is not uniquely defined (see Sec.~\ref{s3}), the
differences of approximate $\epsilon_{\rm x}$'s from each other and from the
``true'' one are not necessarily indicative of a failure of the approximate
functionals.

\begin{figure}[htbp]
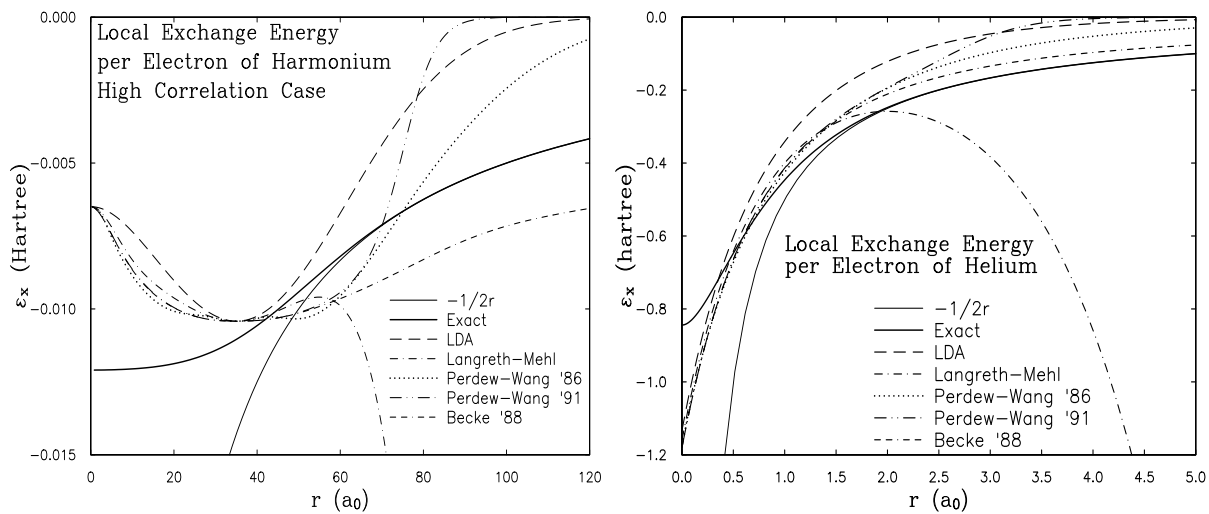

\epsfxsize=8.5 cm
\centerline{{\epsfxsize=7.9 cm \epsfysize=6.7 cm \epsfbox{pl.harm10.ex}}
            {\epsfxsize=7.9 cm \epsfysize=6.7 cm \epsfbox{pl.he.ex}}}
\caption[]{Exact and approximate local exchange energy per electron for
Harmonium ($k\approx 3.6\times 10^{-6}$ a.u.) (left)
and He (right), evaluated for the respective
exact densities.  Taken from Refs.~\cite{harmonium} and \cite{he} and modified.}
\label{harmenx10}
\end{figure}

\begin{figure}[htbp]
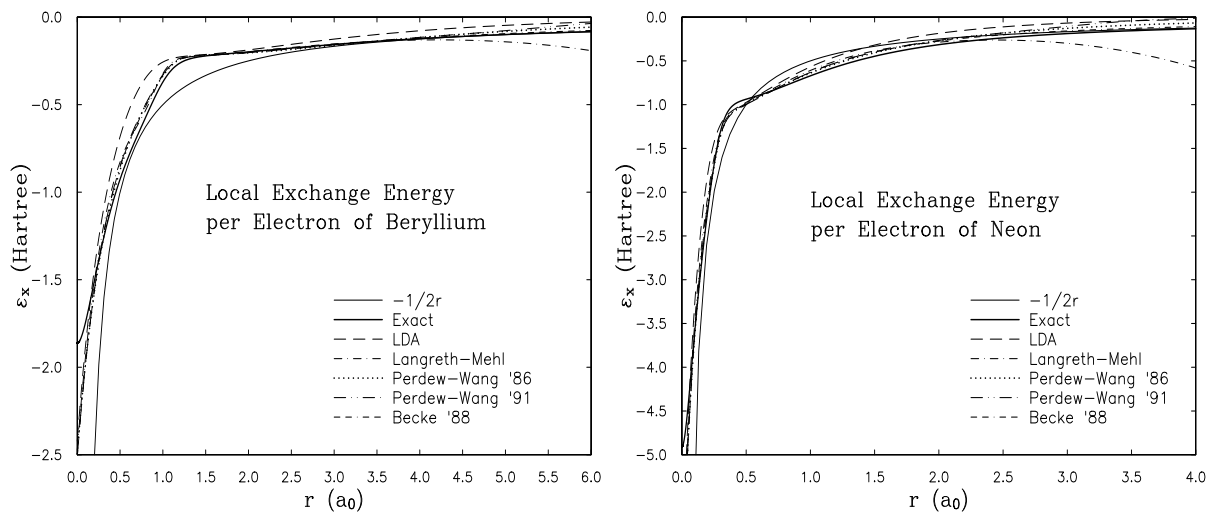

\centerline{{\epsfxsize=7.9 cm \epsfysize=6.7 cm \epsfbox{pl.be.ex}}
            {\epsfxsize=7.9 cm \epsfysize=6.7 cm \epsfbox{pl.ne.ex}}}
\caption[]{Exact and approximate local exchange energy per electron for Be (left)
and Ne (right), evaluated for the respective exact densities.}
\label{bene.ex}
\end{figure}

In the left panel of Fig.~\ref{harmenx10}, we plot the exact and approximate
local exchange energies per electron for Harmonium.
At large {\it r}, the approximate exchange functionals go to zero exponentially
with the exception of the Langreth-Mehl functional which diverges and of the
Becke functional which has the wrong asymptotic behavior $-1/r$ for
Harmonium although it has the correct asymptotic behavior $-1/2r$ for atoms.
The Becke functional was constructed to have the correct asymptotic behavior
for exponentially decaying density but it fails for Gaussian densities (see
Table~\ref{knownprop}).
At the extrema of the density, all the functionals go to the LDA value.

In the right panel of Fig.~\ref{harmenx10} and in Fig.~\ref{bene.ex},
we compare the exact and approximate local exchange
energies per electron of the He, Be and Ne atoms.
At large $r$, the approximate $ \epsilon_{\rm x}$ go to zero exponentially
with the exception of the Langreth-Mehl functional, which diverges, and the
Becke functional which has the correct asymptotic behavior $-1/2r$.
However, as also noted by Engel {\it et~al.}~\cite{ECMV}, the Becke
$\epsilon_{\rm x}$ achieves its asymptotic behavior at much larger distances
than does the true $\epsilon_{\rm x}$.
This is because the correction to the leading $-1/2r$ term is exponentially
small for the true functional but ${\cal O}(1/r^2)$ for the Becke functional.
The Becke exchange energy per electron is however closer to the correct one
over all the range than any other approximate $\epsilon_{\rm x}$.
At intermediate distances, the various GGA's represent an improvement upon LDA,
yielding a deeper $\epsilon_{\rm x}$ than LDA, but, at very short distances,
they are significantly lower than the true one.

\section{COMPARISON OF THE COMPONENTS OF THE TOTAL ENERGY}
\label{s7}

For the model system of two electrons in a harmonic potential, the He
iso-electronic series  and the Be atom, we present a comparison of the
accurate values of $E_{\rm x}$, $E_{\rm c}$ and $E_{\rm xc}$ with the ones
corresponding to approximate functionals evaluated for the accurate densities.
In order to assess the change that results from employing the self-consistent
rather than the exact density, we also perform, for Be, self-consistent
calculations with LDA and the Perdew-Wang '91 GGA and compare the different
components of the total energy (Eq.~\ref{eq0}) with the accurate values.

For the model system of two interacting electrons in a harmonic potential,
we compare in Table~\ref{tab.harm} the values of the exact exchange and
correlation energies with those obtained from the approximate functionals
evaluated for the exact electronic density.
The GGA functionals, except for the Langreth-Mehl functional,
improves considerably upon LDA. The major corrections are achieved for the
exchange contribution. The failure of the Langreth-Mehl functional is a result
of the diverging energy per electron at large $r$. Most of the approximate
functionals are constructed to give accurate values of $E_{\rm xc}$ for atoms
and it is encouraging that they also improve $E_{\rm xc}$ for this rather
different model system.
However, we can see in Tables~\ref{tab.he} and~\ref{tab1.be} that
the improvement upon LDA is much larger for atomic systems: for
instance, the percentage error in the LDA exchange energy evaluated for the
electronic density of He is $14\%$ and in the LDA correlation energy,
$-165\%$, while the percentage errors for the other approximate functionals
we list in the appendix range from $-1$ to $1\%$ for exchange and from
$-3$ to $19\%$ for correlation.
If we consider the sum of the exchange energy and the correlation energy, LDA
yields a good result due to a cancellation of errors and the GGA's
do not improve upon LDA.

\begin{table}[tb]
\caption[]{Model system with spring constant $k\approx 3.6\times 10^{-6}$ a.u.
Exchange-correlation energy for the approximate functionals listed
in Appendix~\ref{a1}. The functionals are evaluated for the
exact density. The exact results are also reported. The numbers in parentheses
are the percentage errors. The total energy of the system is $0.0228$ a.u.
Energies are in Hartree atomic units.
Taken from Ref.~\cite{harmonium} and modified.}
\label{tab.harm}
\begin{tabular*}{\textwidth}{@{}l@{\extracolsep{\fill}}lll}
\hline
Functional        & \multicolumn{1}{c}{$E_{\rm x}$}
                  & \multicolumn{1}{c}{$E_{\rm c}$}
                  & \multicolumn{1}{c}{$E_{\rm xc}$}\\
\hline
Exact KS          &   $-0.0195$      &   $-0.0064$      &   $-0.0259$\\[.1cm]
LDA (Perdew-Wang) &   $-0.0174(11\%)$  &   $-0.0108(-69\%)$ &   $-0.0282(-9\%)$\\
Langreth-Mehl     &   $-0.0217(-11\%)$ &   $-0.0119(-86\%)$ &   $-0.0336(-30\%)$\\
Perdew-Wang '86   &   $-0.0198(-2\%)$  &   $-0.0096(-50\%)$ &   $-0.0294(-14\%)$\\
Perdew-Wang '91   &   $-0.0195(0.2\%)$ &   $-0.0084(-31\%)$ &   $-0.0279(-8\%)$\\
Becke '88         &   $-0.0201(-3\%)$  &   \multicolumn{1}{c}{---}  &  \multicolumn{1}{c}{---} \\
Wilson-Levy       &   \multicolumn{1}{c}{---} &   $-0.0145(-127\%)$ &  \multicolumn{1}{c}{---} \\
Lee-Yang-Parr     &   \multicolumn{1}{c}{---} &   $-0.0032(50\%)$   &  \multicolumn{1}{c}{---} \\
\hline
\end{tabular*}
\end{table}

\begin{sidewaystable}[tbp]
\caption{Comparison of exchange and correlation energies of the He
iso-electronic series for LDA and the various GGA's with the corresponding
exact values. The approximate functionals are evaluated for the exact (not
the self-consistent) densities.
The numbers in parentheses are the percentage errors. All numbers are
accurate to all digits presented with the exception of the ``exact'' results
for Hg$^{+78}$ which probably have an error in the last digit. Taken from
Ref.~\cite{he} and modified.}
\label{tab.he}
\renewcommand{\arraystretch}{1.06}
\renewcommand{\tabcolsep}{.008cm}
\begin{tabular*}{\textwidth}{@{}l@{\extracolsep{\fill}}lrllllllll}
\hline
   & Ion & \multicolumn{1}{c}{Exact}
         & \multicolumn{1}{c}{LDA}
         & \multicolumn{1}{c}{LM}
         & \multicolumn{1}{c}{PW86}
         & \multicolumn{1}{c}{PW91}
         & \multicolumn{1}{c}{B88}
         & \multicolumn{1}{c}{WL}
         & \multicolumn{1}{c}{LYP} \\
   &     &
         & \multicolumn{1}{c}{\cite{PW92}}
         & \multicolumn{1}{c}{\cite{LM83}}
         & \multicolumn{1}{c}{\cite{PW86X,PW86C}}
         & \multicolumn{1}{c}{\cite{PW91}}
         & \multicolumn{1}{c}{\cite{B88}}
         & \multicolumn{1}{c}{\cite{WL}}
         & \multicolumn{1}{c}{\cite{LYP}}\\
\hline
&H$^-$& {\footnotesize$\,\,\,-0.380898$}& {\footnotesize$\,\,\,-0.337(12\%)$}& {\footnotesize$\,\,\,-0.387(-2\%)$}& {\footnotesize$\,\,\,-0.400(-5\%)$}& {\footnotesize$\,\,\,-0.393(-3\%)$}& {\footnotesize$\,\,\,-0.395(-4\%)$}&  \multicolumn{1}{c}{---} &  \multicolumn{1}{c}{---} \\
&He& {\footnotesize$\,\,\,-1.024568$}& {\footnotesize$\,\,\,-0.883(14\%)$}& {\footnotesize$\,\,\,-1.014(1\%)$}& {\footnotesize$\,\,\,-1.033(-1\%)$}& {\footnotesize$\,\,\,-1.016(1\%)$}& {\footnotesize$\,\,\,-1.025(-0\%)$}&  \multicolumn{1}{c}{---} &  \multicolumn{1}{c}{---} \\
E$_{\rm x}$&Be$^{+2}$& {\footnotesize$\,\,\,-2.276628$}& {\footnotesize$\,\,\,-1.957(14\%)$}& {\footnotesize$\,\,\,-2.246(1\%)$}& {\footnotesize$\,\,\,-2.279(-0\%)$}& {\footnotesize$\,\,\,-2.245(1\%)$}& {\footnotesize$\,\,\,-2.265(1\%)$}&  \multicolumn{1}{c}{---} &  \multicolumn{1}{c}{---} \\
&Ne$^{+8}$& {\footnotesize$\,\,\,-6.027484$}& {\footnotesize$\,\,\,-5.173(14\%)$}& {\footnotesize$\,\,\,-5.940(1\%)$}& {\footnotesize$\,\,\,-6.015(0\%)$}& {\footnotesize$\,\,\,-5.927(2\%)$}& {\footnotesize$\,\,\,-5.982(1\%)$}&  \multicolumn{1}{c}{---} &  \multicolumn{1}{c}{---} \\
&Hg$^{+78}$& {\footnotesize$-49.777931$}& {\footnotesize$-42.699(14\%)$}& {\footnotesize$-49.036(1\%)$}& {\footnotesize$-49.605(0\%)$}& {\footnotesize$-48.889(2\%)$}& {\footnotesize$-49.348(1\%)$}&  \multicolumn{1}{c}{---} &  \multicolumn{1}{c}{---} \\[.3cm]
&H$^-$& {\footnotesize$\,\,\,-0.041995$}& {\footnotesize$\,\,\,-0.072(-71\%)$}& {\footnotesize$\,\,\,-0.049(-16\%)$}& {\footnotesize$\,\,\,-0.038(10\%)$}& {\footnotesize$\,\,\,-0.032(24\%)$}& \multicolumn{1}{c}{---} & {\footnotesize$\,\,\,-0.032(25\%)$}& {\footnotesize$\,\,\,-0.030(29\%)$}\\
&He& {\footnotesize$\,\,\,-0.042107$}& {\footnotesize$\,\,\,-0.112(-167\%)$}& {\footnotesize$\,\,\,-0.050(-19\%)$}& {\footnotesize$\,\,\,-0.044(-4\%)$}& {\footnotesize$\,\,\,-0.046(-9\%)$}& \multicolumn{1}{c}{---} & {\footnotesize$\,\,\,-0.042(1\%)$}& {\footnotesize$\,\,\,-0.044(-4\%)$}\\
E$_{\rm c}$&Be$^{+2}$& {\footnotesize$\,\,\,-0.044274$}& {\footnotesize$\,\,\,-0.150(-240\%)$}& {\footnotesize$\,\,\,-0.025(43\%)$}& {\footnotesize$\,\,\,-0.049(-12\%)$}& {\footnotesize$\,\,\,-0.053(-21\%)$}& \multicolumn{1}{c}{---} & {\footnotesize$\,\,\,-0.045(-2\%)$}& {\footnotesize$\,\,\,-0.049(-11\%)$}\\
&Ne$^{+8}$& {\footnotesize$\,\,\,-0.045694$}& {\footnotesize$\,\,\,-0.202(-342\%)$}& {\footnotesize$\;\;\;\;\,0.082(279\%)$}& {\footnotesize$\,\,\,-0.084(-83\%)$}& {\footnotesize$\,\,\,-0.062(-35\%)$}& \multicolumn{1}{c}{---} & {\footnotesize$\,\,\,-0.047(-3\%)$}& {\footnotesize$\,\,\,-0.050(-10\%)$}\\
&Hg$^{+78}$& {\footnotesize$\,\,\,-0.046538$}& {\footnotesize$\,\,\,-0.326(-600\%)$}& {\footnotesize$\;\;\;\;\,1.833(4038\%)$}& {\footnotesize$\,\,\,-0.299(-543\%)$}& {\footnotesize$\,\,\,-0.081(-73\%)$}& \multicolumn{1}{c}{---} & {\footnotesize$\,\,\,-0.048(-3\%)$}& {\footnotesize$\,\,\,-0.051(-9\%)$}\\[.3cm]
&H$^-$& {\footnotesize$\,\,\,-0.422893$}& {\footnotesize$\,\,\,-0.408(3\%)$}& {\footnotesize$\,\,\,-0.435(-3\%)$}& {\footnotesize$\,\,\,-0.438(-4\%)$}& {\footnotesize$\,\,\,-0.425(-0\%)$}& \multicolumn{1}{c}{---} &   \multicolumn{1}{c}{---} &   \multicolumn{1}{c}{---} \\
&He& {\footnotesize$\,\,\,-1.066675$}& {\footnotesize$\,\,\,-0.996(7\%)$}& {\footnotesize$\,\,\,-1.064(0\%)$}& {\footnotesize$\,\,\,-1.077(-1\%)$}& {\footnotesize$\,\,\,-1.062(0\%)$}& \multicolumn{1}{c}{---} &   \multicolumn{1}{c}{---} &   \multicolumn{1}{c}{---} \\
E$_{\rm xc}$&Be$^{+2}$& {\footnotesize$\,\,\,-2.320902$}& {\footnotesize$\,\,\,-2.107(9\%)$}& {\footnotesize$\,\,\,-2.271(2\%)$}& {\footnotesize$\,\,\,-2.328(-0\%)$}& {\footnotesize$\,\,\,-2.298(1\%)$}& \multicolumn{1}{c}{---} &   \multicolumn{1}{c}{---} &   \multicolumn{1}{c}{---} \\
&Ne$^{+8}$& {\footnotesize$\,\,\,-6.073178$}& {\footnotesize$\,\,\,-5.375(11\%)$}& {\footnotesize$\,\,\,-5.858(4\%)$}& {\footnotesize$\,\,\,-6.099(-0\%)$}& {\footnotesize$\,\,\,-5.989(1\%)$}& \multicolumn{1}{c}{---} &   \multicolumn{1}{c}{---} &   \multicolumn{1}{c}{---} \\
&Hg$^{+78}$& {\footnotesize$-49.824470$}& {\footnotesize$-43.025(14\%)$}& {\footnotesize$-47.203(5\%)$}& {\footnotesize$-49.904(-0\%)$}& {\footnotesize$-48.970(2\%)$}& \multicolumn{1}{c}{---} &   \multicolumn{1}{c}{---} &   \multicolumn{1}{c}{---}\\
\end{tabular*}
\end{sidewaystable}

In Tables~\ref{tab.he} and~\ref{tab1.be}, we compare the approximate
$E_{\rm x}$, $E_{\rm c}$ and $E_{\rm xc}$, evaluated for the accurate densities
of the He iso-electronic series and Be respectively, with the corresponding accurate quantities.
As is well known, LDA yields values of $E_{\rm x}$ that are too small in
absolute magnitude and values of $E_{\rm c}$ that are too large in absolute
magnitude, resulting in a cancellation of errors for $E_{\rm xc}$.
All the GGA's yield improved exchange, correlation and exchange-correlation
energies compared to the LDA.

\begin{table}[tbp]
\catcode`?=\active \def?{\kern\digitwidth}
\caption{Comparison of exchange and correlation energies of Be for LDA and the
various GGA's with the corresponding accurate values. The approximate
functionals are evaluated for the exact (not the self-consistent) densities.
The numbers in parentheses are the percentage errors. Energies are in Hartree
atomic units.  Taken from Ref.~\cite{UG}.}
\label{tab1.be}
\begin{tabular*}{\textwidth}{@{}l@{\extracolsep{\fill}}lll}
\hline
Functional        & \multicolumn{1}{c}{$E_{\rm x}$}
                  & \multicolumn{1}{c}{$E_{\rm c}$}
                  & \multicolumn{1}{c}{$E_{\rm xc}$}\\
\hline
Accurate
             & $-2.67398$     &  $-0.09615$       & $-2.77013$ \\[.1cm]
LDA (Perdew-Wang)
             & $-2.32125(13\%)$ &  $-0.22515(-134\%)$ & $-2.54641(8\%)$\\
Langreth-Mehl
             & $-2.60576(3\%)$  &  $-0.10012(-4\%)$   & $-2.70588(2\%)$\\
Perdew-Wang '86
             & $-2.69050(-1\%)$ &  $-0.09425(2\%)$    & $-2.78475(-1\%)$\\
Perdew-Wang '91
             & $-2.65434(1\%)$  &  $-0.09496(1\%)$    & $-2.74929(1\%)$\\
Becke '88
             & $-2.66694(0\%)$  & \multicolumn{1}{c}{---}  &\multicolumn{1}{c}{---}\\
Wilson-Levy
             & \multicolumn{1}{c}{---}  &  $-0.09572(0\%)$ &\multicolumn{1}{c}{---}\\
Lee-Yang-Parr
             & \multicolumn{1}{c}{---}  &  $-0.09546(1\%)$ &\multicolumn{1}{c}{---}\\
\hline
\end{tabular*}
\end{table}

\begin{table}[tbp]
\newlength{\digitwidth} \settowidth{\digitwidth}{\rm 0}
\catcode`?=\active \def?{\kern\digitwidth}
\caption[]{Comparison of LDA, Perdew-Wang '91 GGA and accurate values for
various components of the Kohn-Sham total energy, the density-weighted integral
of $v_{\rm xc}$ and single-particle eigenvalues of Be.
The approximate functionals are evaluated for the respective self-consistent
densities.
The accurate values probably have errors in the last digit quoted.
The accurate value of the total energy is from Ref.~\cite{atom}.
Energies are in Hartree atomic units.  Taken from Ref.~\cite{UG}.}
\label{tab2.be}
\begin{tabular*}{\textwidth}{@{}l@{\extracolsep{\fill}}rrrrr}
\hline
Property & \multicolumn{1}{c}{LDA}
         & \multicolumn{1}{c}{PW91}
         & \multicolumn{1}{c}{Accurate}
         & \multicolumn{1}{c}{$\Delta_{\rm LDA}$}
         & \multicolumn{1}{c}{$\Delta_{\rm PW91}$}\\
\hline
$E=T_{\rm s}+E_{\rm en}+E_{\rm H}+E_{\rm xc}$
     & $-14.44647$ & $-14.64794$ & $-14.66736$ &  $0.22089$ &  $0.01942$\\
$T_{\rm s}$
     &  $14.30879$ & $14.57461$  & $14.5942?$  & $-0.2854?$ & $-0.0196?$ \\
$E_{\rm en}$
     & $-33.35610$ & $-33.65670$ & $-33.7099?$ &  $0.3538?$ &  $0.0532?$ \\
$E_{\rm H}$
     &   $7.11487$ &   $7.17261$ &   $7.2185?$ & $-0.1036?$ & $-0.0459?$ \\
$E_{\rm xc}$
     &  $-2.51403$ &  $-2.73845$ &  $-2.77013$ &  $0.25610$ &  $0.03168$\\[.3cm]
$E_{\rm x}$
     &  $-2.29033$ &  $-2.64428$ &  $-2.67398$ &  $0.38365$ &  $0.02970$\\
$E_{\rm c}$
     &  $-0.22370$ &  $-0.09417$ &  $-0.09615$ & $-0.12754$ &  $0.00198$\\[.3cm]
$E_{\rm en}+E_{\rm H}$
     & $-26.24123$ & $-26.48410$ & $-26.4914?$ &  $0.2502?$ &  $0.0073?$\\[.3cm]
$\int{\rm d}{\bf r}\;\rho({\bf r})\;v_{\rm xc}({\bf r})$
     &  $-3.30614$ &  $-3.50058$ &  $-4.4595?$ &  $1.1533?$ &  $0.9589?$ \\
$\epsilon_{1s}$
     &  $-3.85609$ &  $-3.91154$ &  $-4.2265?$ &  $0.3704?$ &  $0.3149?$ \\
$\epsilon_{2s}$
     &  $-0.20577$ &  $-0.20719$ &  $-0.3426?$ &  $0.1368?$ &  $0.1354?$\\
\hline
\end{tabular*}
\end{table}
For the He iso-electronic series, the true correlation energies are nearly
constant, whereas for the Be iso-electronic series there is a linear component
in the nuclear charge $Z$.
It is
difficult for any GGA functional to mimic the correct $Z$ dependence of both
iso-electronic series.

In Table~\ref{tab2.be}, we show various quantities, computed by LDA and the
Perdew-Wang '91 GGA, and the corresponding accurate values for Be. LDA and
GGA values were obtained through a self-consistent procedure and are
therefore evaluated for the LDA and the GGA density respectively.
The change in $E_{\rm x}$ and $E_{\rm c}$ due to self-consistency are not
negligible.
The values of $E_{\rm x}$ and $E_{\rm c}$ in Table~\ref{tab2.be} differ from
those in Table~\ref{tab2.be} by a larger amount for LDA than for the
Perdew-Wang '91 GGA reflecting our earlier observation that the GGA yields
somewhat more accurate self-consistent densities than does LDA.
The last two columns show the errors in LDA and GGA.
The LDA components of the total energy are all smaller in absolute magnitude
than the corresponding accurate values. Both the total energy and its
components have much smaller errors in GGA than in LDA.
In Fig.~\ref{be.den}, we observed that, near the nucleus, the LDA density is
too small while GGA gives an increased charge density due to the
lower value at short distances and the negative divergence at the nucleus
of the GGA exchange-correlation potential. This increase in the GGA density
results in an increase in the magnitude of all the components of the total
energy.
Consequently, the GGA yields improvements not only in $E_{\rm xc}$ but
also in other components of the total energy, despite, or may be one should
say because of, the spurious divergence of the exchange-correlation
at the nucleus.
On the other hand, quantities related to the single-particle eigenvalues
(shown in the last two rows) are not improved in GGA relative to
LDA. In particular, the highest lying eigenvalue which, for the true density
functional, should equal the ionization energy, is not much improved.
This is to be expected since the long-distance behavior of
exchange-correlation potential, which strongly influences the highest
eigenvalue, is not appreciably improved.
All the observations made here for the Be atom are also true for the
Ne atom (see Refs.~\cite{proc} and \cite{UG}).

\section{PROSPECTS}
\label{s8}

Although the exchange and correlation energies calculated from the
various GGA's (with the exception of the Langreth-Mehl GGA for the model
system and for the correlation energies in the He iso-electronic series)
are considerably more accurate than those from LDA, the situation is not
as clear for the exchange and correlation potentials.
Since $E_{\rm x}$ and $E_{\rm c}$ are integrated quantities, it is possible
for GGA's to improve upon them without making comparable improvements to
the exchange and correlation potentials which contain more detailed
information.

The various GGA exchange functionals have certain features in common.
By favoring regions with large density gradients, they correctly lead to more
negative total energies and reduced binding energies relative to LDA.
They yield a lower potential than LDA at short and intermediate distances
from nuclei and approximately reproduce the intershell structure -- two steps
in the right direction -- but they also introduce a spurious divergence at
the nucleus and fail to improve the long-range behavior.
Whereas the various GGA exchange functionals have some common features,
the GGA correlation functionals are very different from each other and none
of them yields a correlation potential resembling the true one. Correlation
is more subtle than exchange as manifested in several ways.
1) The exchange potential is everywhere negative, whereas the correlation
potential can be either negative or positive in different regions of space.
2) For He iso-electronic series, the correlation potential varies less over
space for heavy ions than for light atoms~\cite{he}. This is opposite
to the behavior in LDA.
3) The correlation potential has qualitatively the same behavior for all members
of the helium iso-electronic series~\cite{he} (negative at short distances
and long distance, positive at intermediate distances, see Fig.~\ref{vxc.he})
but it has the opposite qualitative behavior for the closely related model
system of two electrons in a harmonic potential (Fig.~\ref{vxc10}).

As discussed in Sec.~\ref{s1}, any GGA that includes no higher than first
derivatives of the density cannot simultaneously satisfy both the correct
$-1/r$ behavior of the exchange-correlation potential and the correct
$-1/2r$ behavior of exchange energy per electron at large
distances. Moreover, any such GGA must have a spurious divergence in the
exchange-correlation potential at nuclei.
We mentioned that, by including the Laplacian of the density in an appropriate
way, it is possible to construct a GGA that has the correct long-range
asymptotics of the exchange potential and energy per electron and
the proper short-range behavior of the exchange-correlation potential.
Nevertheless, it may turn out that even GGA functionals containing Laplacian
terms are not flexible enough to achieve a major improvement and it may be
necessary to consider more general functionals.
We discuss here some possibilities.

Since a data-base on accurate exchange-correlation potentials has being
built in recent years and none of the existing GGA functionals passes
the test of the comparison with the true potentials, more effort is now being
devoted to developing schemes that focus on reproducing the main features
of the true exchange-correlation potential.
As mentioned above, the potential controls the quality of the self-consistent
density and is the key ingredient in evaluating energy gradients~\cite{LB2},
\begin{eqnarray}
\frac{\partial E_{\rm xc}\left[\rho\right]}{\partial {\rm X}}=
\int v_{\rm xc}\left([\rho];{\bf r}\right)\frac{\partial \rho
\left({\bf r}\right)} {\partial {\rm X}} \,{\rm d}{\bf r},
\end{eqnarray}
whose accurate estimation is for instance necessary in determining potential
energy surfaces.

Instead of starting from an exchange-correlation energy functional and then
deriving the corresponding potential, some authors start with an expression
for the potential which has the correct asymptotic behavior~\cite{LB1,LRC}.
While it is easy to construct a potential with the correct asymptotics,
one encounters the problem of how to obtain the exchange-correlation energy
functional that corresponds to this potential.
In fact, the potential may not be the functional derivative of an energy
expression.
If the approximate exchange potential satisfies the uniform scaling property
$v_{\rm x}\left(\left[\rho_\lambda\right];{\bf r}\right)=
\lambda v_{\rm x}\left(\left[\rho\right];\lambda{\bf r}\right)$, a
possibility is to use the Levy-Perdew relation to assign a corresponding
exchange energy~\cite{LP85}:
\begin{eqnarray}
E_{\rm x}=\int v_{\rm x}\left([\rho];{\bf r}\right)\left(3\rho({\bf r})
+{\bf r}\cdot\nabla\rho({\bf r})\right){\rm d}{\bf r}.\label{rel}
\end{eqnarray}
The relation is derived by considering a line integral of the potential along
the density path $\rho_\lambda$ (Eq.~\ref{scale}) and is exact only for a
potential that is a functional derivative and has the correct scaling property:
\begin{eqnarray}
E_{\rm x}\left[\rho\right]=\int_0^1 {\rm d}\lambda\int{\rm d}{\bf r}\;
v_{\rm x}\left(\left[\rho_\lambda\right];{\bf r}\right)
\frac{{\rm d}\rho_\lambda({\bf r})}{{\rm d}\lambda}
=\int v_{\rm x}\left([\rho];{\bf r}\right)\left(3\rho({\bf r})
+{\bf r}\cdot\nabla\rho({\bf r})\right){\rm d}{\bf r}.
\end{eqnarray}
The first equality always holds for a potential that is a functional
derivative while the second one is valid only if the potential has the
correct behavior under uniform scaling.
Van Leeuwen and Baerends~\cite{LB2} discuss why Eq.~\ref{rel} is a sensible
choice for a potential that is not a functional derivative but has the proper
scaling property and is constructed to mimic the true potential at the given
density. Unfortunately, such a relation does not exist in the case of exchange
and correlation since the scaling properties of the correlation potential are
not homogeneous~\cite{LP93}.
Assigning an exchange-correlation energy to a given potential becomes more
complicated because we need to know the potential not only for the density
of interest, as in the case of exchange-only, but along some path in density
space~\cite{LB2}.

Another attempt to generate a DFT scheme yielding a better description of
the ex\-change-correlation potential has as a starting point the OEP method.
The OEP method provides the exact solution for the problem of exchange-only
DFT but, as shown by Norman and Koelling~\cite{NK}, it can be
generalized to treat exchange and correlation if the correlation energy
is also expressed as a functional of the Kohn-Sham orbitals.
The resulting problem is technically difficult to solve but Krieger, Li and
Iafrate proposed an approximation that greatly simplify the approach~\cite{KLI}.
Although the resulting exchange-correlation potential is no longer equivalent
to the functional derivative with respect to the density of the original
exchange-correlation energy, it closely corresponds to the exact generalized
OEP exchange-correlation potential if the expression
$\delta E_{\rm xc}\left[\{\psi_i\}\right]/
\delta\psi_i({\bf r})/ \psi_i^*({\bf r})$ goes to zero at large distances.
Since the true exchange energy is known as a functional of the Kohn-Sham 
orbitals (Eq.~\ref{enx1}) and can therefore be employed, the approximate
exchange potential, evaluated according to the procedure of Ref.~\cite{KLI},
has the correct short and long-distance asymptotic behavior although 
the intershell bump is underestimated. An improved approximation that
yields a more accurate intershell bump is also discussed in Ref.~\cite{KLI}.
This approach is of practical interest since it is easier to construct
self-interaction free, exchange-correlation functionals that depend on the
orbitals instead of the density.
Grabo and Gross~\cite{GG} recently implemented this scheme with the
approximation of Krieger, Li and Iafrate and the orbital-dependent correlation
functional of Colle and Salvetti~\cite{CS}.

A different approximate scheme that cures the long-distance asymptotics of the
ex\-change-correlation potential is obtained by combining a given approximate
exchange-correlation functional with a self-interaction correction (SIC)
procedure~\cite{SIC}. A GGA depending on the Laplacian of the density could be
easily constructed so that the exchange-correlation potential does not have a
spurious divergence at nuclei and could then be implemented in a SIC scheme
to yield a potential with also the correct long-range asymptotic behavior.

\section*{Acknowledgments}
We thank John Morgan and Jonathan Baker for making available their program
for calculating very accurate wavefunctions for the He iso-electronic series,
Manfred Taut for providing the program for calculating the densities of
Harmonium and John Perdew the routine for calculating the Perdew-Wang 91
functional.
We benefited from several useful discussions with Mel Levy and Eberhard Gross.
The calculations were performed on the IBM SP2 computer at the Cornell Theory
Center.
This work is supported by the Office of Naval Research and NATO (grant number
CRG 940594).
X.G. acknowledges financial support from FNRS-Belgium and the
European Union (Human Capital and Mobility Program contract CHRX-CT940462).

\appendix
\section{APPROXIMATE FUNCTIONAL FORMS OF {\large $E_{\rm xc}\left[\rho\right]$}}
\label{a1}

The second order generalized gradient approximation exchange-correlation
energy is written as
\begin{eqnarray}
E_{\rm xc}^{\rm GGA}\left[\rho\right]=\int {\rm d}{\bf r}
       \,e_{\rm xc}(\rho({\bf r}),\left|\nabla \rho({\bf r})\right|,
       \nabla^2 \rho({\bf r})).
\end{eqnarray}
We define the following density dependent variables:
\begin{eqnarray*}
k_{\rm F}=\left(3\pi^2\rho\right)^{1/3},\;\;
k_s=\left(\frac{4}{\pi}k_{\rm F}\right)^{1/2},
\;\;s=\frac{\left|\nabla \rho\right|}{2\,k_{\rm F}\,\rho},\;\;
t=\frac{\left|\nabla \rho\right|}{2 k_s\,\rho},\;\;
r_s=\left(\frac{3}{4\pi\rho}\right)^{1/3}.
\end{eqnarray*}
We present the spin unpolarized version of the approximate functionals,
$\zeta=(\rho_\uparrow-\rho_\downarrow)/\rho=0$.
All the parameters that appear in the following functionals are in atomic
units.
\subsection*{LDA exchange functional}
\begin{eqnarray}
e^{\rm LDA}_{\rm x}=A_x\rho^{4/3}\,,
\end{eqnarray}
where $A_{\rm x}=-(3/4)\left(3/\pi\right)^{1/3}$.
\subsection*{LDA correlation functional (Perdew-Wang \cite{PW92})}
\begin{eqnarray}
e^{\rm LDA}_{\rm c}=- 2a\rho\left(1+\alpha_1 r_s\right)
\log\left[1+\frac{1}{2 a(\beta_1 r_s^{1/2}+\beta_2 r_s+\beta_3 r_s^{3/2}
+\beta_4 r_s^{2})}\right]\,,
\end{eqnarray}
where $a=0.0310907$, $\alpha_1=0.21370$, $\beta_1=7.5957$, $\beta_2=3.5876$,
$\beta_3=1.6382$ and $\beta_4=0.49294$.
\subsection*{RPA correlation functional \cite{BL}}
\begin{eqnarray}
e_{\rm c}&=&-\rho\,c^{\rm P} f(r_s/r^{\rm p}),\\
f(z)&=&(1+z^3)\ln\left(1+\frac{1}{z}\right)+\frac{z}{2}-z^2-\frac{1}{3}\,,
\nonumber
\end{eqnarray}
where $c^{\rm P}=0.0252$ and $r^{\rm P}=30$.

\subsection*{Langreth-Mehl exchange-correlation functional \cite{LM83}}
\begin{eqnarray}
e_{\rm x}&=&e^{\rm LDA}_{\rm x}-a\frac{\left|\nabla \rho\right|^2}{\rho^{4/3}}
  \left(\frac{7}{9}+18\,f^2\right)\,,\\
e_{\rm c}&=&e^{\rm RPA}_{\rm c}\left(\rho\right)
 +a\,\frac{\left|\nabla \rho\right|^2}{\rho^{4/3}}\left(2e^{-F}+18f^2\right)\,,
\end{eqnarray}
where $F=b\left|\nabla \rho\right|/\rho^{7/6}$, $b=\left(9\pi\right)^{1/6}f$,
$a=\pi/(16(3\pi^2)^{4/3})$ and $f=0.15$.
\subsection*{Perdew-Wang '86 exchange functional \cite{PW86X}}
\begin{eqnarray}
e_{\rm x}=e_{\rm x}^{\rm LDA}(\rho)\left(1+0.0864\,\frac{s^2}{m}+b\,s^4
+c\,s^6\right)^{m}\,,
\end{eqnarray}
where $m=1/15$, $b=14$ and $c=0.2$.
\subsection*{Perdew-Wang '86 correlation functional \cite{PW86C}}
\begin{eqnarray}
e_{\rm c}=e_{\rm c}^{\rm LDA}(\rho)+e^{-\Phi}
C_{\rm c}(\rho)\frac{\left|\nabla \rho\right|^2}{\rho^{4/3}}\,,
\end{eqnarray}
where
\begin{eqnarray}
\Phi&=&1.745\,\tilde{f}\,\frac{C_{\rm c}(\rho=\infty)}{C_{\rm c}(\rho)}\,
\frac{\left|\nabla \rho\right|} {\rho^{7/6}}\,,\nonumber\\
C_{\rm c}(\rho)&=&C_1+\frac{C_2+C_3 r_s+C_4 r_s^2}
{1+C_5 r_s +C_6 r_s^2 +C_7 r_s^3}\nonumber
\end{eqnarray}
and $\tilde{f}=0.11$, $C_1=0.001667$, $C_2=0.002568$, $C_3=0.023266$,
$C_4=7.389\cdot 10^{-6}$, $C_5=8.723$, $C_6=0.472$, $C_7=7.389\cdot 10^{-2}$.
\subsection*{Perdew-Wang '91 exchange functional \cite{PW91}}
\begin{eqnarray}
e_{\rm x}=e_{\rm x}^{\rm LDA}(\rho)\left[
\frac{1+a_1\,s\,{\rm sinh}^{-1}(a_2 s)+\left(a_3+a_4 e^{-100\,s^2}\right)\,s^2}
          {1+a_1\,s\,{\rm sinh}^{-1}(a_2 s)+a_5 s^4}\right]\,,
\end{eqnarray}
where $a_1=0.19645$, $a_2=7.7956$, $a_3=0.2743$, $a_4=-0.1508$ and $a_5=0.004$.
\subsection*{Perdew-Wang '91 correlation functional \cite{PW91}}
\begin{eqnarray}
e_{\rm c}=\left[e^{LDA}_{\rm c}(\rho)+\rho\,H(\rho,s,t)\right]\,,
\end{eqnarray}
where
\begin{eqnarray}
H&=&\frac{\beta^2}{2\alpha}\log\left[1+\frac{2\alpha}{\beta}
\frac{t^2+A\,t^4}{1+A\,t^2+A^2t^4}\right]+
C_{c0}\left[C_{\rm c}(\rho)-C_{c1}\right]t^2 e^{-100\,s^2}\,,\nonumber\\
A&=&\frac{2\alpha}{\beta}\left[e^{-2\alpha\epsilon_{\rm c}^{\rm LDA}(\rho)/
\beta^2}-1\right]^{-1} \nonumber
\end{eqnarray}
and $\alpha=0.09$, $\beta=0.0667263212$, $C_{c0}=15.7559$, $C_{c1}=0.003521$.
The function $C_{\rm c}(\rho)$ is the same as for the Perdew-Wang '86
correlation functional.
$\epsilon_{\rm c}^{\rm LDA}(\rho)=e^{LDA}_{\rm c}(\rho)/\rho$.

%
%
%

\subsection*{Becke '88 exchange functional \cite{B88}}

\begin{eqnarray}
e_{\rm x}=e_{\rm x}^{\rm LDA}(\rho)\left[1-\frac{\beta}{2^{1/3}A_{\rm x}}
\frac{x^2}{1+6\beta\,x\,{\rm sinh}^{-1}(x)}\right]\,,
\end{eqnarray}
where $x=2\left(6\pi^2\right)^{1/3}s=2^{1/3}\left|\nabla\,\rho\right|/
\rho^{4/3}$, $A_{\rm x}=-(3/4)\left(3/\pi\right)^{1/3}$ and
$\beta=0.0042$.

\subsection*{Engel-Chevary-Macdonald-Vosko exchange functional \cite{ECMV}}

\begin{eqnarray}
e_{\rm x}=e_{\rm x}^{\rm LDA}(\rho)\frac{1+a_1\xi+a_2\xi^2}
{1+b_1\xi+b_2\xi^2}\,,
\end{eqnarray}
where $\xi=s^2$ and $a_1=27.8428$, $a_2=11.7683$, $b_1=27.5026$ and
$b_2=5.7728$.

\subsection*{Wilson-Levy correlation functional \cite{WL}}
\begin{eqnarray}
e_{\rm c}=\frac{a\,\rho+b\left|\nabla \rho\right|/\rho^{1/3}}
   {c+d\left|\nabla \rho\right|/(\rho/2)^{4/3}+r_s}\,,
\end{eqnarray}
where $a=-0.74860$, $b=0.06001$, $c=3.60073$ and $d=0.90000$.
\subsection*{Closed shell Lee-Yang-Parr correlation functional \cite{LYP}}
\begin{eqnarray}
 e_{\rm c}=-a\frac{1}{1+d\,\rho^{-1/3}}\left\{\,\rho
+b\,\rho^{-2/3}\left[C_F\,\rho^{5/3}- 2t_W+\frac{1}{9}
\left(t_W+\frac{1}{2}\nabla^2 \rho\right)\right]e^{-c\,\rho^{-1/3}}\right\}\,,
\end{eqnarray}
where
\begin{eqnarray}
t_W=\frac{1}{8}\left(\frac{\left|\nabla
\rho\right|^2}{\rho}-\nabla^2\,\rho\right)\nonumber
\end{eqnarray}
and $C_F=3/10\left(3\pi^2\right)^{2/3}$, $a=0.04918$, $b=0.132$, $c=0.2533$
and $d=0.349$.


\begin{thebibliography}{99}
\bibitem{HK} P. Hohenberg and W. Kohn, Phys. Rev. {\bf 136}, B864 (1964).
\bibitem{KS} W. Kohn and L. J. Sham, Phys. Rev. {\bf 140}, A1133 (1976).
\bibitem{JG} R. O. Jones and O. Gunnarsson, Rev. Mod. Phys. {\bf 61}, 689
(1989).
\bibitem{WKK} C. S. Wang, B. M. Klein and H. Krakauer, Phys. Rev. Lett.
              {\bf 54}, 1852 (1985).
\bibitem{Herman69} F. Herman, J. P. van Dyke and I. B. Ortenburger,
Phys. Rev. Lett. {\bf 22}, 807 (1969); F. Herman, I. B. Ortenburger and J. P.
Van Dyke, Int. J. Quantum Chem. Symp. {\bf 3}, 827 (1970).
\bibitem{PK} A. E. DePristo and J. D. Kress, J. Chem. Phys. {\bf 86}, 1425
(1987).
\bibitem{MB} S. -K. Ma and K. A. Brueckner, Phys. Rev. {\bf 165}, 19 (1968).
\bibitem{PLS} J. P. Perdew, D. C. Langreth and V. Sahni, Phys. Rev. Lett.
{\bf 38}, 1030 (1977).
\bibitem{LM83} D. C. Langreth and M. J. Mehl, Phys. Rev. B {\bf 28}, 1809
(1983).
\bibitem{Perdew85} J. P. Perdew, Phys. Rev. Lett. {\bf 55}, 1665 (1985).
\bibitem{B88} A. D. Becke, Phys. Rev. A {\bf 33}, 3098 (1988).
\bibitem{PW86X} J. P. Perdew and Y. Wang, Phys. Rev. B {\bf 33}, 8800 (1986).
\bibitem{PW86C} J. P. Perdew, Phys. Rev. B {\bf 33}, 8822 (1986); erratum
{\it ibid}. {\bf 34}, 7406 (1986).
\bibitem{PW91}   J. P. Perdew in {\it Electronic structure of solids '91},
edited by P. Ziesche and H. Eschrig (Akademie Verlag, Berlin, 1991);
J. P. Perdew, K. Burke and Y. Wang, submitted to Phys. Rev. B.
\bibitem{PW92} J. P. Perdew and Y. Wang, Phys. Rev. B {\bf 45}, 13244 (1992).
\bibitem{WL} L. C. Wilson and M. Levy, Phys. Rev. B {\bf 41}, 12930 (1990).
\bibitem{LYP} C. Lee, W. Yang and R. G. Parr, Phys. Rev. B {\bf 37}, 785 (1988).
\bibitem{ECMV} E. Engel, J. A. Chevary, L. D. Macdonald and S. H. Vosko,
Z. Phys. D {\bf 23}, 7 (1992).
\bibitem{SIC} J. P. Perdew and A. Zunger, Phys. Rev. B {\bf 23}, 5048 (1981);
H. Stoll, C. M. E. Pavlidou and H. Preuss, Theor. Chim. Acta {\bf 49}, 143 
(1978); S. H. Vosko and L. Wilk, J. Phys. B {\bf 16}, 3687 (1983).
\bibitem{OEP} R. T. Sharp and G. K. Horton, Phys. Rev. {\bf 90}, 317 (1953);
J. D. Talman, and W. F. Shadwick, Phys. Rev. A {\bf 14}, 36 (1976).
\bibitem{GJL} O. Gunnarsson, M. Jonson and B. I. Lundqvist, Phys. Rev. B
{\bf 20}, 3136 (1979).
\bibitem{GJ} O. Gunnarsson and R. O. Jones, Phys. Scr. {\bf 21}, 394 (1980).
\bibitem{P} J. P. Perdew, J. A. Chevary, S. H. Vosko, K. A. Jackson,
M. R. Pederson, D. J. Singh and C. Fiolhhais, Phys. Rev. B {\bf 46}, 6671
(1992).
\bibitem{KP} F. W. Kutzler and G. S. Painter, Phys. Rev. B {\bf 45}, 3236 
(1992).
\bibitem{B92} A. D. Becke, J. Chem. Phys. {\bf 96}, 2155 (1992); {\it ibid.}
{\bf 97}, 9173 (1992).
\bibitem{JGP} B. G. Johnson, P. M. W. Gill and J. A. Pople, J. Chem. Phys.
{\bf 98}, 5612 (1993).
\bibitem{OB} G. Ortiz and P. Ballone, Phys. Rev. B {\bf 43}, 6376 (1991).
\bibitem{WATER} K. Laasonen, F. Csajka and M. Parrinello, Chem. Phys. Lett.
{\bf 194}, 172 (1992); C. Lee, D. Vanderbildt, K. Laasonen, R. Car and M.
Parrinello, Phys. Rev. B {\bf 47}, 4863 (1993).
\bibitem{BJG} P. Bagno, O. Jepsen and O. Gunnarsson, Phys. Rev. B {\bf 40},
1997 (1989).
\bibitem{semicond} C. Filippi, D. J. Singh and C. J. Umrigar, Phys. Rev. B
{\bf 50}, 19947 (1994), and references therein.  
\bibitem{metal} A. Khein, D. J. Singh and C. J. Umrigar, Phys. Rev. B {\bf 51},
4105 (1995), and references therein.
\bibitem{S79} D. W. Smith, S. Jagannathan, and G. S. Handler,
Int. J. Quantum Chem. Symp. {\bf 13}, 103 (1979).
\bibitem{vonBarth84} U. von Barth in {\it Electronic Structure of Complex
Systems}, edited by P. Phariseau and  W. M. Temmerman, NATO, ASI Series B,
vol. 113 (1984).
\bibitem{AP} C. -O. Almbladh and A. C. Pedroza, Phys. Rev. B {\bf 29}, 2322
(1984); A. C. Pedroza, Phys. Rev. A  {\bf 33}, 804 (1986).
\bibitem{AS} F. Aryasetiawan and M. J. Stott, Phys. Rev. B {\bf 34}, 4401
(1986); {\it ibid.} {\bf 38}, 2974 (1988).
\bibitem{NM} A. Nagy and N. H. March, Phys. Rev. A {\bf 39}, 5512 (1989);
{\it ibid.} {\bf 40}, 554 (1989).
\bibitem{D90} E. R. Davidson, Int. J. Quant. Chem. {\bf 37}, 811 (1990).
\bibitem{Chen94} J. Chen, R. O. Esquivel, and M. J. Stott,
Philos. Mag. B {\bf 69}, 1001 (1994).
\bibitem{LB1} R. van Leeuwen and E. J. Baerends, Phys. Rev. A {\bf 49}, 2421
(1994).
\bibitem{MZ} R. C. Morrison and Q. Zhao, Phys. Rev. A {\bf 51}, 1980 (1995).
\bibitem{Ludena95} E. V. Lude\~na, J. Maldonado, R. L\'opez-Boada,
T. Koga, and E. S. Kryachko, J. Chem. Phys. {\bf 102}, 318 (1995).
\bibitem{BBS} M. A. Buijse, E. J. Baerends and J. G. Snijders, Phys. Rev. A
{\bf 40}, 4190 (1989).
\bibitem{GLB} O. V. Gritsenko, R. van Leeuwen and E. J. Baerends, Phys. Rev. A
{\bf 52}, 1870 (1995).
\bibitem{IH} V. E. Ingamells and N. C. Handy, Chem. Phys. Lett. {\bf 248},
373 (1996).
\bibitem{proc} C. J. Umrigar and X. Gonze, in {\it High Performance Computing
and its Application to the Physical Sciences}, proceedings of the Mardi Gras
'93 Conference, edited by D. A. Browne {\it et~al.}, (World Scientific,
Singapore, 1993).
\bibitem{KG} W. Knorr and R. W. Godby, Phys. Rev. Lett. {\bf 68}, 639 (1992).
\bibitem{UG} C. J. Umrigar and X. Gonze, unpublished (1996).
\bibitem{GLxc} O. Gunnarsson and B. I. Lundqvist, Phys. Rev. B {\bf 13}, 4274
(1976)
\bibitem{LP93} M. Levy and J. P. Perdew, Phys. Rev. B {\bf 48}, 11638 (1993).
\bibitem{LP85} M. Levy and J. P. Perdew,  Phys. Rev. A  {\bf 32}, 2010 (1985).
\bibitem{L91} M. Levy, Phys. Rev. A {\bf 43}, 4637 (1991).
\bibitem{GLs} A. G\"orling and M. Levy, Phys. Rev. A {\bf 45}, 1509 (1992).
\bibitem{LPr} M. Levy and J. P. Perdew, Int. J. Quant. Chem. {\bf 49}, 539
(1993), and references therein.
\bibitem{harmonium} C. Filippi, C. J. Umrigar and M. Taut,  J. Chem. Phys.
{\bf 100}, 1290 (1994).
\bibitem{he} C. J. Umrigar and X. Gonze, Phys. Rev. A {\bf 50}, 3827 (1994).
\bibitem{U_notes} C. J. Umrigar, unpublished notes (1993).
\bibitem{JK} P. \"Jemmer and P. J. Knowles, Phys. Rev. A {\bf 51}, 3571 (1995).
\bibitem{EV2} E. Engel and S. H. Vosko, Phys. Rev. B {\bf 50}, 10498 (1994).
\bibitem{M} N. H. March, Phys. Rev. A {\bf 36}, 5077 (1987).
\bibitem{SGNB} P. S\"ule, O. V. Gritsenko, A. Nagy and E. J. Baerends,
J. Chem. Phys. {\bf 103}, 10085 (1995).
\bibitem{GL} A. G\"orling and M. Levy, Int. J. Quant. Chem, Symp. {\bf 29},
93 (1995); Phys. Rev. A {\bf 50}, 196 (1994).
\bibitem{CCX} C. Filippi, C. J. Umrigar and X. Gonze, submitted to
Phys. Rev. A.
\bibitem{T}  M. Taut, Phys. Rev. A {\bf 48}, 3561 (1993).
\bibitem{Morgan} The wavefunction used is a minor modification of that in
D. E. Freund, B. D. Huxtable and J. D. Morgan,  Phys. Rev. A  {\bf 29}, 980 (1984).
\bibitem{AB} C. -O. Almbladh and U. von Barth,  Phys. Rev. B  {\bf 31}, 3231
(1985).
\bibitem{UNR} C. J. Umrigar, P. Nightingale and K. J. Runge, J. Chem. Phys.
{\bf 99}, 2865 (1993).
\bibitem{gfmcbook} D. M. Ceperley and M. H. Kalos in {\it Monte Carlo Methods
in Statistical Physics} edited by K. Binder, Topics Current Phys., Vol.7
(Springer, Berlin, Heidelberg, 1979).
\bibitem{Patil} S. H. Patil, J. Phys. B {\bf 23}, 1 (1990).
\bibitem{EV1} E. Engel and S. H. Vosko, Phys. Rev. B {\bf 47}, 13164 (1993).
\bibitem{atom} E. R. Davidson, S. A. Hagstrom, S. J. Chakravorty, V. M. Umar
and C. F. Fischer, Phys. Rev. A {\bf 44}, 7071 (1991); S. J. Chakravorty,
S. R. Gwaltney, E. R. Davidson, F. A. Parpia and C. F. Fischer, Phys. Rev. A
{\bf 47}, 3649 (1993).
\bibitem{LB2} R. van Leeuwen and E. J. Baerends, Phys. Rev. A {\bf 51}, 170
(1995).
\bibitem{NK} M. R. Norman and D. D. Koelling, Phys. Rev. B {\bf 30}, 5530
(1984).
\bibitem{LRC} A. Lembarki, F. Rogemond and H. Chermette, Phys. Rev. A {\bf 52},
3704 (1995).
\bibitem{KLI} J. B. Krieger, Y. Li and G. J. Iafrate, Phys. Rev. A {\bf 46},
5453 (1992).
\bibitem{GG} T. Grabo and E. K. U. Gross, Chem. Phys. Lett. {\bf 240}, 141
(1995).
\bibitem{CS} R. Colle and D. Salvetti, Theoret. Chim. Acta {\bf 37}, 329
(1975); {\it ibid.} {\bf 53}, 55 (1979).
\bibitem{BL}  U. von Barth and L. Hedin, J. Phys. C {\bf 5}, 1629 (1972).

\end{thebibliography}
\end{document}